\documentclass{nature}
\usepackage{graphicx}
\usepackage{amsmath}
\usepackage{amssymb}

\title{Formation of secondary atmospheres on terrestrial planets by late disk accretion}

\author{Quentin Kral$^{1}$, Jeanne Davoult$^1$ \& Benjamin Charnay$^1$}

\begin{document}

\maketitle

\begin{affiliations}
 \item LESIA, Observatoire de Paris, Université PSL, CNRS, Sorbonne Université, Univ. Paris Diderot, Sorbonne Paris Cité, 5 place Jules Janssen, 92195 Meudon, France
\end{affiliations}

\begin{abstract}
Recently, gas disks have been discovered around main sequence stars well beyond the usual protoplanetary disk lifetimes (i.e., $\gtrsim$ 10 Myrs), when planets have already formed\cite{2014Sci...343.1490D,2016ApJ...828...25L,2017ApJ...849..123M,2019AJ....157..117M}. These gas disks, mainly composed of CO, carbon, and oxygen\cite{2014A&A...563A..66C,2017MNRAS.469..521K,2017ApJ...839L..14H} seem to be ubiquitous\cite{2017ApJ...849..123M} in systems with planetesimal belts (similar to our Kuiper belt), and can last for hundreds of millions of years\cite{2017ApJ...842....9M}. Planets orbiting in these gas disks will accrete\cite{2014ApJ...797...95L,2015ApJ...811...41L} a large quantity of gas that will transform their primordial atmospheres into new secondary atmospheres with compositions similar to that of the parent gas disk. Here, we quantify how large a secondary atmosphere can be created for a variety of observed gas disks and for a wide range of planet types. We find that gas accretion in this late phase is very significant and an Earth's atmospheric mass of gas is readily accreted on terrestrial planets in very tenuous gas disks. In slightly more massive disks, we show that massive CO atmospheres can be accreted, forming planets with up to sub-Neptune-like pressures. Our new results demonstrate that new secondary atmospheres with high metallicities and high C/O ratios will be created in these late gas disks, resetting their primordial compositions inherited from the protoplanetary disk phase, and providing a new birth to planets that lost their atmosphere to photoevaporation or giant impacts. We therefore propose a new paradigm for the formation of atmospheres on low-mass planets, which can be tested with future observations (JWST, ELT, ARIEL). We also show that this late accretion would show a very clear signature in Sub-Neptunes or cold exo-Jupiters. Finally, we find that accretion creates cavities in late gas disks, which could be used as a new planet detection method, for low mass planets a few au to a few tens of au from their host stars.
\end{abstract}

The discovery of large amounts of gas around main-sequence stars is recent with most detections occurring in the last few years\cite{2013ApJ...776...77K,2017ApJ...849..123M}. These late gas disks are observed in systems that have planetesimal belts, which are older than 10 Myr and can last for hundreds of millions of years\cite{2017ApJ...842....9M}. It is thought that the observed gas is released from volatile-rich planetesimals when they collide with each other in the system's belts\cite{2012ApJ...758...77Z,2017MNRAS.469..521K}. The gas then viscously evolves\cite{2016MNRAS.461.1614K}, spreading inward and outwards\cite{2016MNRAS.461..845K,2018arXiv181108439K}. Hence the observed gas is likely secondary (rather than of primordial origin) and this late disk (main-sequence) phase is different from the younger ($<$10 Myr) protoplanetary disks that are much more massive, hydrogen-rich and in which giant planets form within a few millions of years\cite{2015ApJ...812L..38T}.

These late gas disks are nearly ubiquitous around A-type stars; Gas has been detected around more than 70\% of systems with bright planetesimal belts\cite{2017ApJ...849..123M}. As for other stellar type stars or lower mass systems the statistics are still based on too small a sample as these gas disks are harder to detect but gas evolution models\cite{2017MNRAS.469..521K} predict that all stars surrounded by planetesimal belts should have gas at a certain level that depends on the mass of solids in the system's planetesimal belt. 
More than 25\% of stars have planetesimal belts massive enough to be detected through their infrared-excess\cite{2014MNRAS.445.2558T}, and it may be that most stars have belts below current detection limits (for instance an exact equivalent of the Kuiper-belt around a nearby star is not massive enough to be detectable with current facilities). These late gas disks might therefore not be the exception but rather the rule.

At $>$10 Myr in these late gas disks, planets have already formed. For instance, we now observe massive planets as early as 5 Myr\cite{2018A&A...617A..44K, 2018A&A...617L...2M,2019NatAs.tmp..329H} and we know that terrestrial planets such as Mars formed very early ($<$10 Myr from cosmochemical evidence\cite{2007Icar..191..497N}) and the Earth took slightly longer (10-30 Myr) but most of its mass was acquired within 10 Myr\cite{2005AREPS..33..531J}. The planets embedded in these disks will be able to accrete the disk gas in a similar way as when planets accrete gas in younger protoplanetary disks but for much longer timescales because late gas disks can last for tens of millions of years. In this paper, we estimate how much gas can be accreted in this late phase onto the already-formed planets, and whether this gas can create new secondary atmospheres (with a composition similar to the source gas disk, i.e. rich in carbon and oxygen and depleted in hydrogen) that would replace their original atmospheres.

To compute the amount of gas accretion onto terrestrial or Super-Earth planets in these late disks, we used an accretion model first developed for protoplanetary disk environments\cite{2014ApJ...797...95L,2015ApJ...811...41L} that we adapted to work in the late disk phase studied in this paper (Methods). A planet embedded in a gas disk will quickly fill its Hill sphere and whether it is able to accrete mass from the Hill sphere to its atmosphere depends on how fast the planet can cool (radiate away energy), and then contract to eventually accrete some more gas into its Hill sphere\cite{2015ApJ...811...41L} and grow an atmosphere. In the protoplanetary disk case, it is predicted that gas accretion should not depend strongly on disk density. However, in late gas disks (much more tenuous than protoplanetary disks), the mass available can be much lower than the mass that can be accreted from cooling/contraction and atmosphere growth is therefore limited by the gas disk mass (Methods).

\begin{figure}
    \includegraphics[width=11cm]{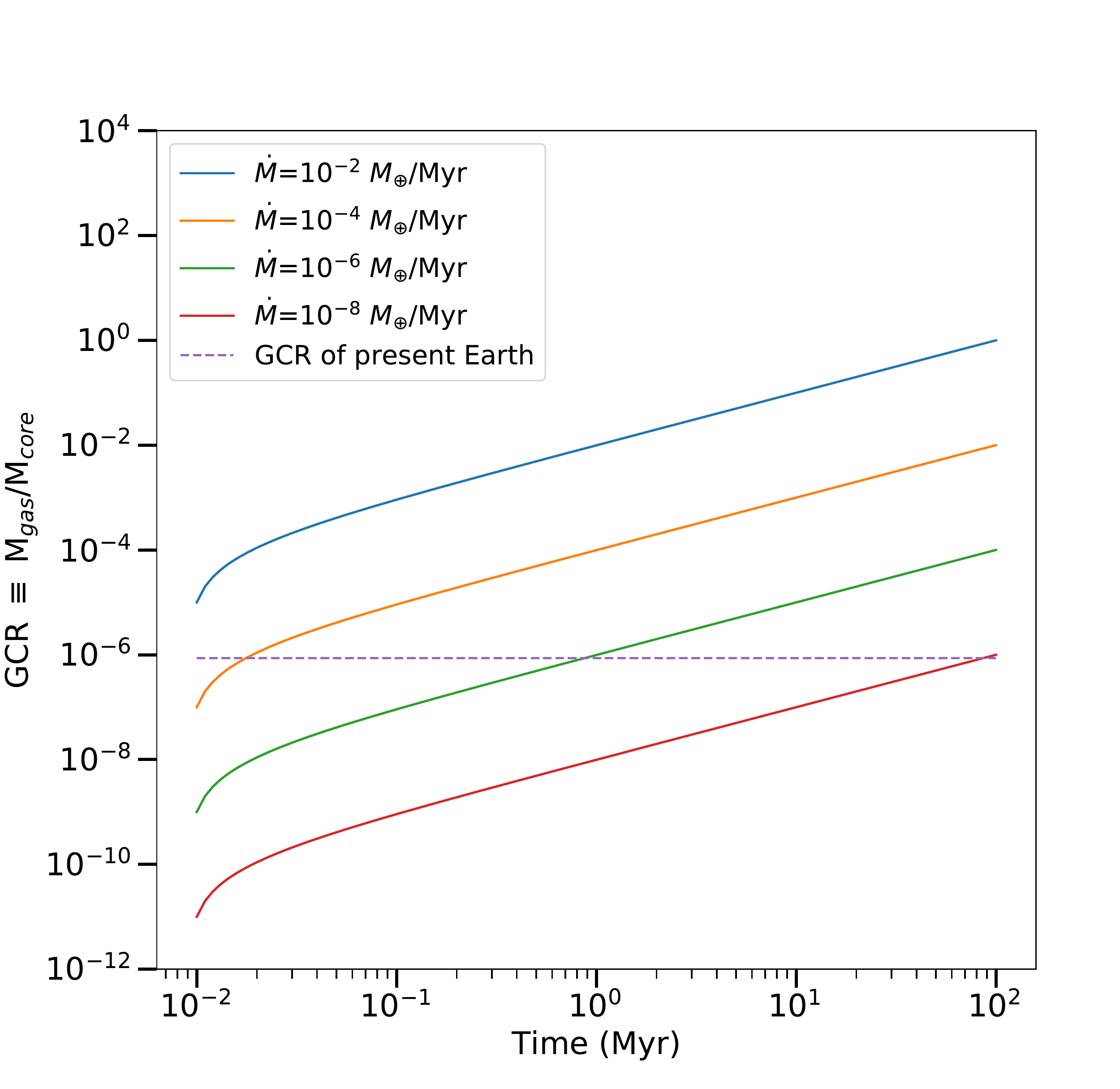}
\caption{{\bf Formation of massive secondary atmospheres.} Temporal evolution of the gas-to-core ratio (GCR) of an Earth-like (1 M$_\oplus$ at 1 au) planet starting with no atmosphere and orbiting in a late gas disk, for disks with gas crossing rates $\dot{M}$ at the planet varying from $10^{-8}$ and $10^{-2}$ M$_\oplus$/Myr. The dashed line shows the Earth's GCR at $8.6\times 10^{-7}$.}
\label{figgcr}

\end{figure}

We run the accretion model starting from an Earth-like planet of mass 1 M$_\oplus$ at 1 au from its host star and assuming that the planet's atmosphere has no, or very little, mass, because of photoevaporation (i.e. it is in the desiccated part of the radius valley)\cite{2019AREPS..47...67O} or desiccated after a large impact\cite{2019MNRAS.486.2780Y} (that should happen frequently in the late stages of planetary formation\cite{2016ApJ...821..126Q}). Figure~\ref{figgcr} shows the temporal evolution of the gas-to-core ratio (GCR) of the accreting planet for different gas crossing rates from the disk $\dot{M}$ at the planet location (a more massive belt provides higher $\dot{M}$, see methods). If the disk is at steady state, $\dot{M}$ is the same as the gas release rate in the planetesimal belt\cite{2016MNRAS.461..845K}. For the most massive belts, we observe that the CO input rate in the belt can reach $\sim10^{-1}$ M$_\oplus$/Myr but for less massive belts releasing less CO per unit time, we expect values that can go below $10^{-7}$ M$_\oplus$/Myr\cite{2017ApJ...842....9M} and this is why we explore the effect of different $\dot{M}$ ($10^{-8}$, $10^{-6}$, $10^{-4}$, $10^{-2}$ M$_\oplus$/Myr) on the total atmospheric mass accreted by a planet. We assume that most of the accretion happens within 100 Myr (Methods). Mass accretion for different accretion times can be straightforwardly extrapolated from Figure~\ref{figgcr}.

\begin{figure}
\centering
   \includegraphics[width=8cm]{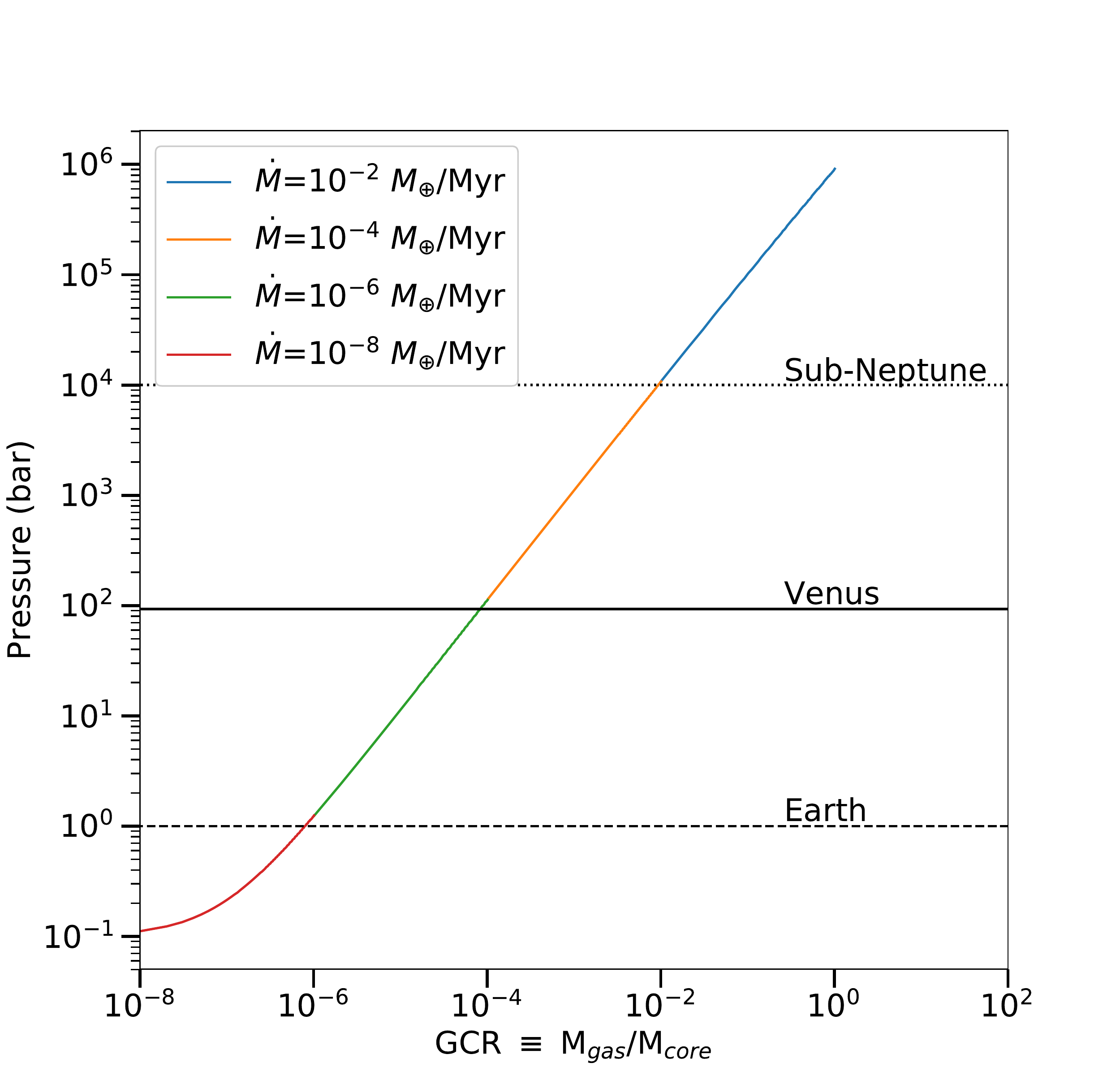}
    \includegraphics[width=8cm]{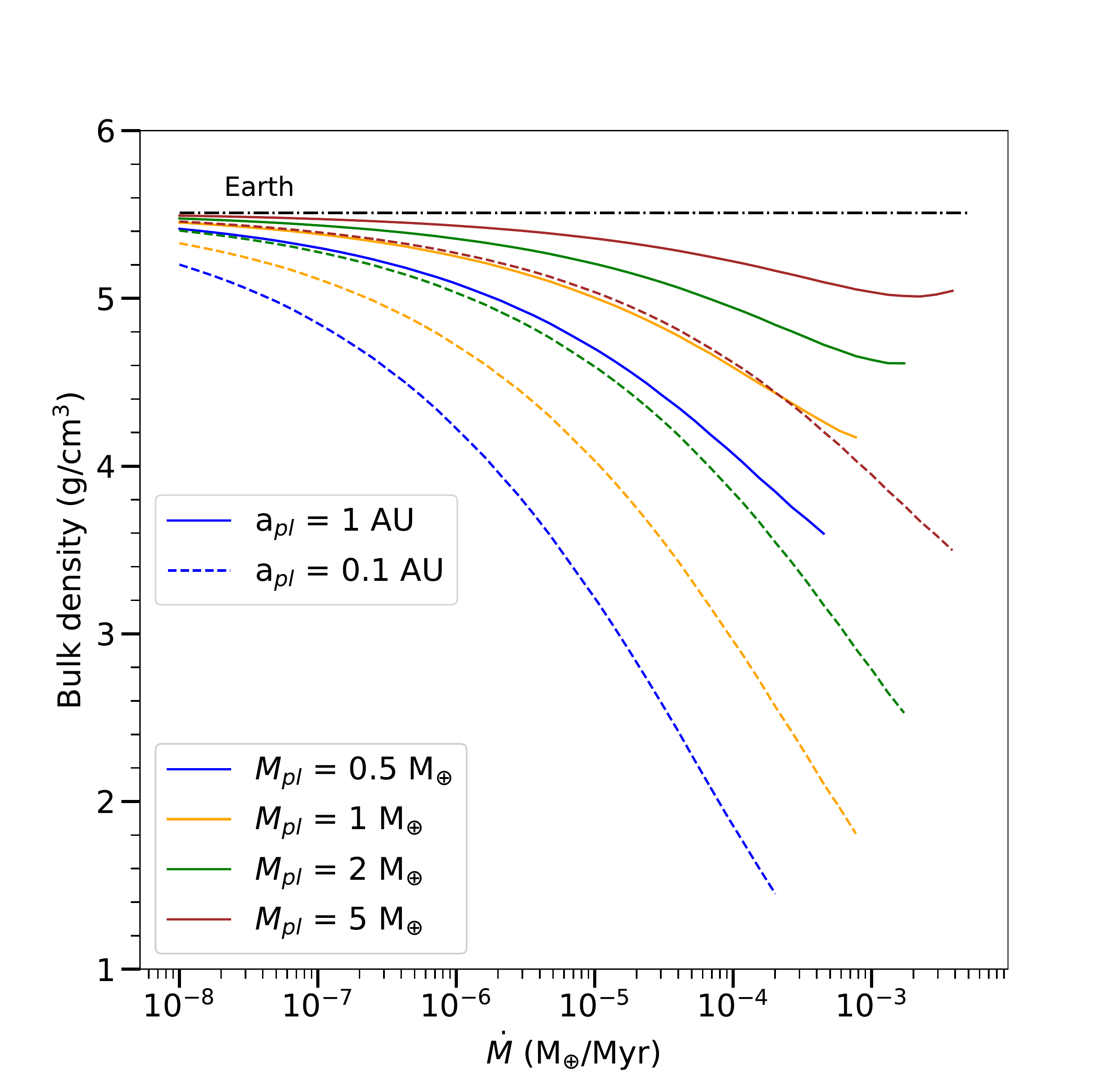}
\caption{{\bf Pressure and Density evolution of an initially desiccated planet embedded in a late gas disk.} {\it Left:} Pressure Vs. GCR of an Earth-like (1 M$_\oplus$ at 1 au) planet starting with no atmosphere for different gas input rates, evolving up to 100 Myr. {\it Right:} Bulk density Vs. $\dot{M}$ for different planet masses and semi-major axes, plotted at 100 Myr up to a maximum GCR of 0.1.}
\label{figdens}

\end{figure}

We find that a planet starting with a low atmospheric mass will very rapidly accrete gas even at low gas input rates $\dot{M}$. Indeed, within 1 Myr, all simulations with $\dot{M} \geq 10^{-6}$ M$_\oplus$/Myr accreted more than an Earth's atmosphere mass. Given enough time, late accretion transforms a terrestrial planet (1 M$_\oplus$ at 1 au) with no atmospheres (or even starting with an Earth-like or Venus atmosphere, see Extended data Figure~\ref{figpre}) into a planet with a massive atmosphere with a GCR up to $>10^{-2}$ for $\dot{M} \geq 10^{-4}$ M$_\oplus$/Myr. We also show that this late accretion is very efficient on both larger (e.g., 5 M$_\oplus$ super-Earth) and smaller (e.g., Mars-size) planets and for both closer (e.g., at 0.1 au) and more distant planets (e.g., at 10 au) in Fig.~\ref{figdens} and in Extended data Figure~\ref{figmars}. We note that in our accretion scenario, there is no risk of runaway because the gas the planet can accrete from is limited and creating a Jupiter from a terrestrial mass core in these low-mass disks is simply not possible (Methods).

In Figure~\ref{figdens}, we show the effect of gas accretion on planet's pressure and bulk density over time (Methods) and confirm that a variety of pressures from Earth-like (1 bar) to Sub-Neptune-like ($>10^{4}$ bar) can be reached on low-mass planets formed in these disks (Fig.~\ref{figdens} left). Even though pressures as high as $10^5$ bar (Neptune-like) can be reached on these planets, their densities never reach values as low as that of Neptune because CO atmospheres have a much higher mean molecular weight compared to hydrogen-dominated atmospheres. We note that in the new accretion scenario proposed here, CO accretion does not depend much on the core mass (at least for planets with $R_H/H>1$, Methods) so that the density is higher for higher mass cores. We thus predict statistically that rocky planets (with $R_H/H>1$ and excluding H$_2$-dominated atmospheres) with larger cores will have higher densities, which is in contrast with current models of planet formation and may be in line with current observations not seeing any clear correlation between the core mass and GCR but rather a large spread in GCRs\cite{2014ApJ...792....1L,2016ApJ...817...90L}. We also note that as accretion is very efficient, we expect the outermost planets to accrete most disk material before it has time to spread further in, leading to decreasing densities for increasing semi-major axes (for a given core mass but note that for planets with $R_H/H<1$ accretion will be slightly less efficient, Methods). There is a competing but less important effect, where closer-in planets have higher temperatures and thus lower densities for a given accreted mass (Fig.~\ref{figdens} right). For instance, our scenario may explain systems such as TRAPPIST-1 with decreasing densities ($5.64\pm0.4$,$4.50\pm0.20$,$4.18\pm0.19$,$3.96\pm0.6$)\cite{2018A&A...613A..68G} for the 4 outermost planets (e, f, g and h, respectively), where the low densities of TRAPPIST-1 f, g and h could be explained by massive CO atmospheres of $\sim 10^5$ bar.

Another strong prediction for this type of accretion is that planet atmospheres formed with this mechanism should have a high mean molecular weight and be mostly made of carbon and oxygen (rather than hydrogen) with a C/O ratio close to 1 (as CO is expected to be the main volatile released but see Methods for detail as other molecules may also be released). Current HST observations of super-Earths revealed flat transit spectra, interpreted as the presence of atmospheres with high mean molecular weights and clouds\cite{2014Natur.505...69K,2015ApJ...813L...1C,2015ApJ...815..110M}.
JWST and ARIEL will have the power (Methods) to take spectra for many more terrestrial planets (e.g. the temperate TRAPPIST-1 planets\cite{2017ApJ...850..121M}) and Super-Earths and find out whether it is common for these planets to have such high metallicities (and high C/O ratio) to test whether this new phase of late accretion really happens widely. 

We find that our late accretion scenario is much more efficient and favourable to accrete volatiles on a terrestrial planet than delivery from impacts, i.e., even if a large heavy bombardment-like event happens after several 100s of Myr after the gas disk dissipated, the atmosphere would still be dominated by volatiles accreted by late gas accretion (Supplementary information). The impact and late gas accretion scenarios could be distinguished based on the final composition of the observed planets, in particular looking at their C/O ratio (Supplementary information). We also find that our scenario of late accretion works for a wide range of planets going from Mars-like to Super-earths and for close-in as well as distant planets from their stars. It also works if the planet is not initially devoid of atmosphere and has an Earth-like or a Venus atmosphere initially as it would replace the bulk of these atmospheres with gas coming from late disk accretion, even for cases where the disk is not very massive (Methods). Finally, if the gas accretion happens only for 10 Myr or for a period longer than 100 Myr, we find that the quantity of gas accreted by planets always remains considerable leading to planetary atmospheres with masses at least that of the Earth's atmosphere for disks more massive than typical very-low-mass disks such as the Kuiper belt (Methods).

\begin{figure}
\centering
    	\includegraphics[width=8cm]{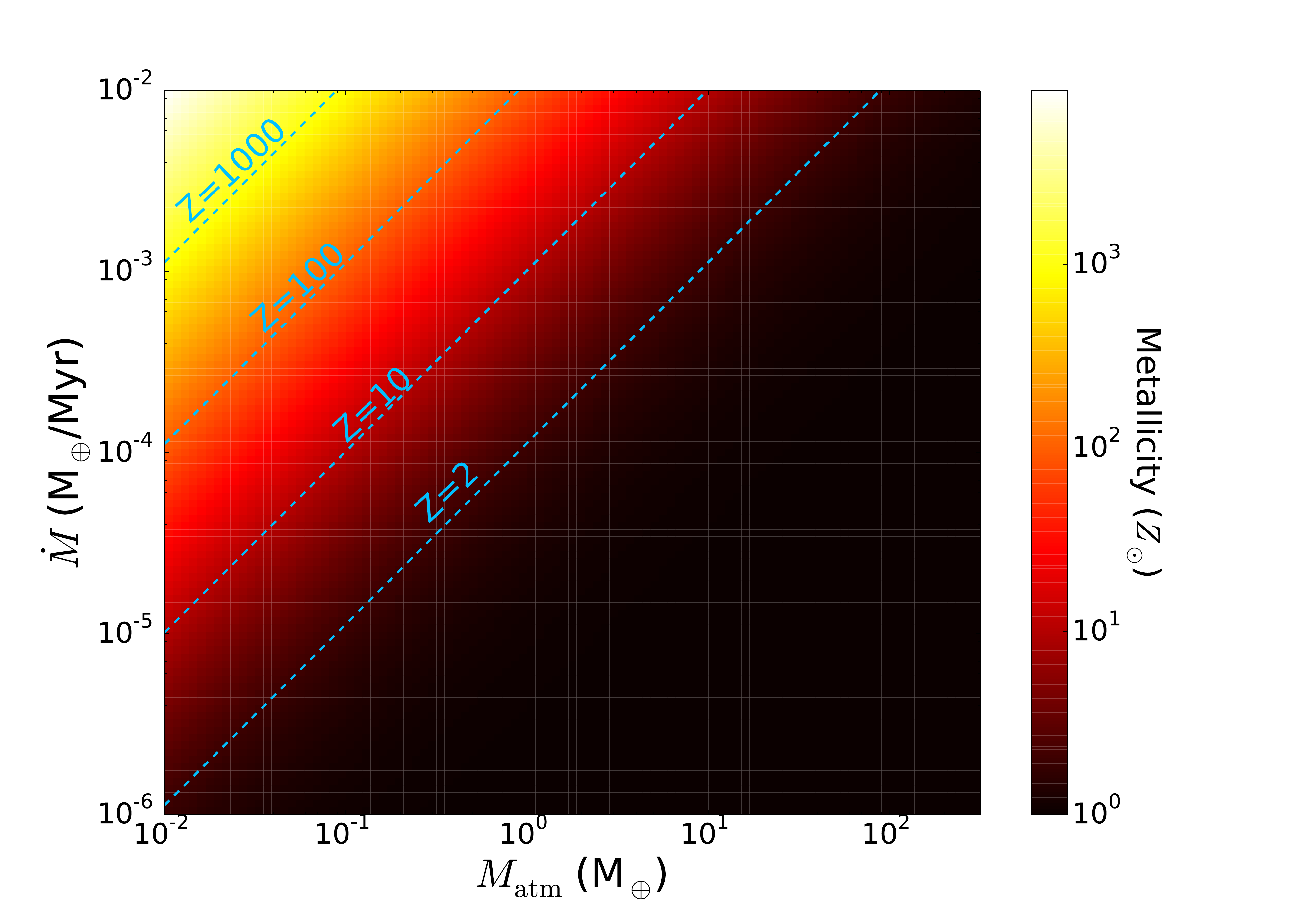}
        \includegraphics[width=8cm]{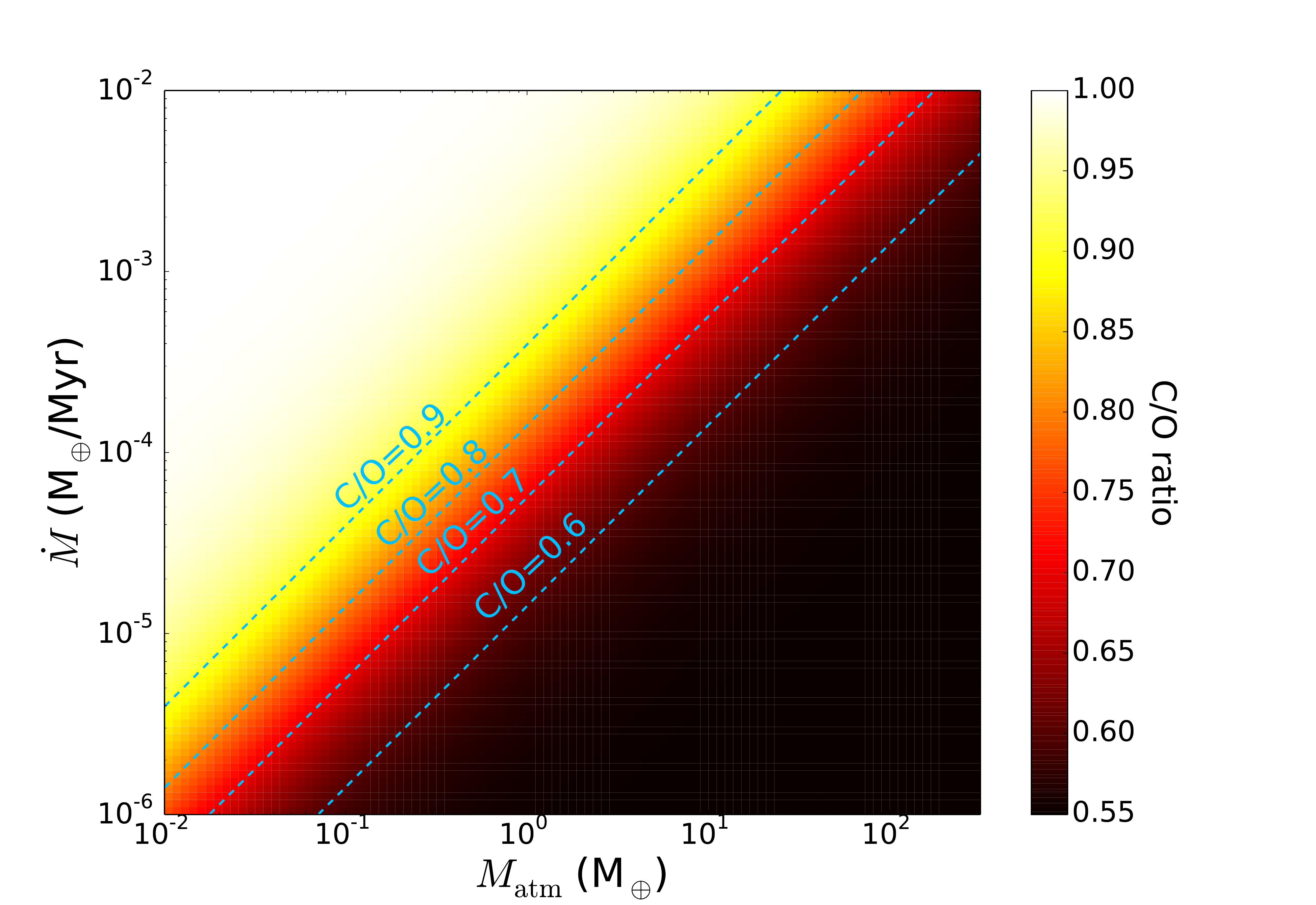}

\caption{{\bf Signature of late gas accretion on giant planets.} Temporal variation of metallicity (left) and C/O ratio (right) as accretion proceeds for 100 Myr from an initially hydrogen-rich primordial atmosphere for different gas input rates and different initial atmosphere masses (Sub-Neptune up to Jupiter).}
\label{figmet100}

\end{figure}

We also show that for existing (initially hydrogen-rich) Sub-Neptunes or more massive planets, the accretion is also going to affect their atmospheres once they mix with this new gas (Supplementary information). In Figure~\ref{figmet100} (left), we show that the metallicity in Sub-Neptunes could reach $>1000$ times the solar metallicity, down to a factor 1.25 for more massive Jupiter-like planets. There are now some direct measurements of atmospheric metallicities in Neptunes and Sub-Neptunes\cite{2014Natur.513..526F,2019NatAs...3..813B}. Some studies find near-solar metallicities (e.g., GJ 3470 b)\cite{2019NatAs...3..813B}, while some others find super-solar metallicities (e.g., HAT-P-11 b or HAT-P-26 b or K2-18 b)\cite{2014Natur.513..526F, 2017Sci...356..628W, 2019ApJ...887L..14B, 2019NatAs...3.1086T} and more measurements will be welcome to test our scenario. The C/O ratio in these giant planets may also become close to 1 for a great variety of atmosphere masses and $\dot{M}$, and even increase by 10\% in a Jupiter-like planet for large values of $\dot{M}$ (Figure~\ref{figmet100} right). In Supplementary data Figure~2, we plot the metallicity and C/O ratio after 10 Myr of evolution to show that these effects can occur early in the planet's history. Measurements of C/O ratios are still scarce but first results show super-solar (e.g. HR 8799 c)\cite{2013Sci...339.1398K} as well as sub-solar values (e.g. $\beta$ Pic b)\cite{2019arXiv191204651G} and more measurements, especially for smaller Sub-Neptune planets, will help to test our scenario.

We also make predictions for detecting ongoing accretion onto young Jupiter-like planets or more massive brown dwarfs in direct imaging (Supplementary information). When the gas accretes onto the outer envelope of the giant planet, it accumulates and diffuses inwards over time\cite{2018ApJ...854..172C}. We show that this accumulation (see Supplementary data figure~3) will be detectable on spectra observed with JWST-NIRCam or JWST-NIRSpec as well as instruments on ELTs in the M-band around 4.5-5 $\mu$m (see Supplementary data figure~4) for the coldest giant planets or brown dwarfs ($<$800 K), which would be a clear signature of this accretion. Only a few spectra of planets cooler than 800 K (e.g. GJ 504b\cite{2013ApJ...778L...4J}) have been obtained so far in direct imaging but none targeted the required CO bands\cite{2015Sci...350...64M,2016ApJ...817..166S}. 

\begin{figure}
\centering
    \includegraphics[width=11cm]{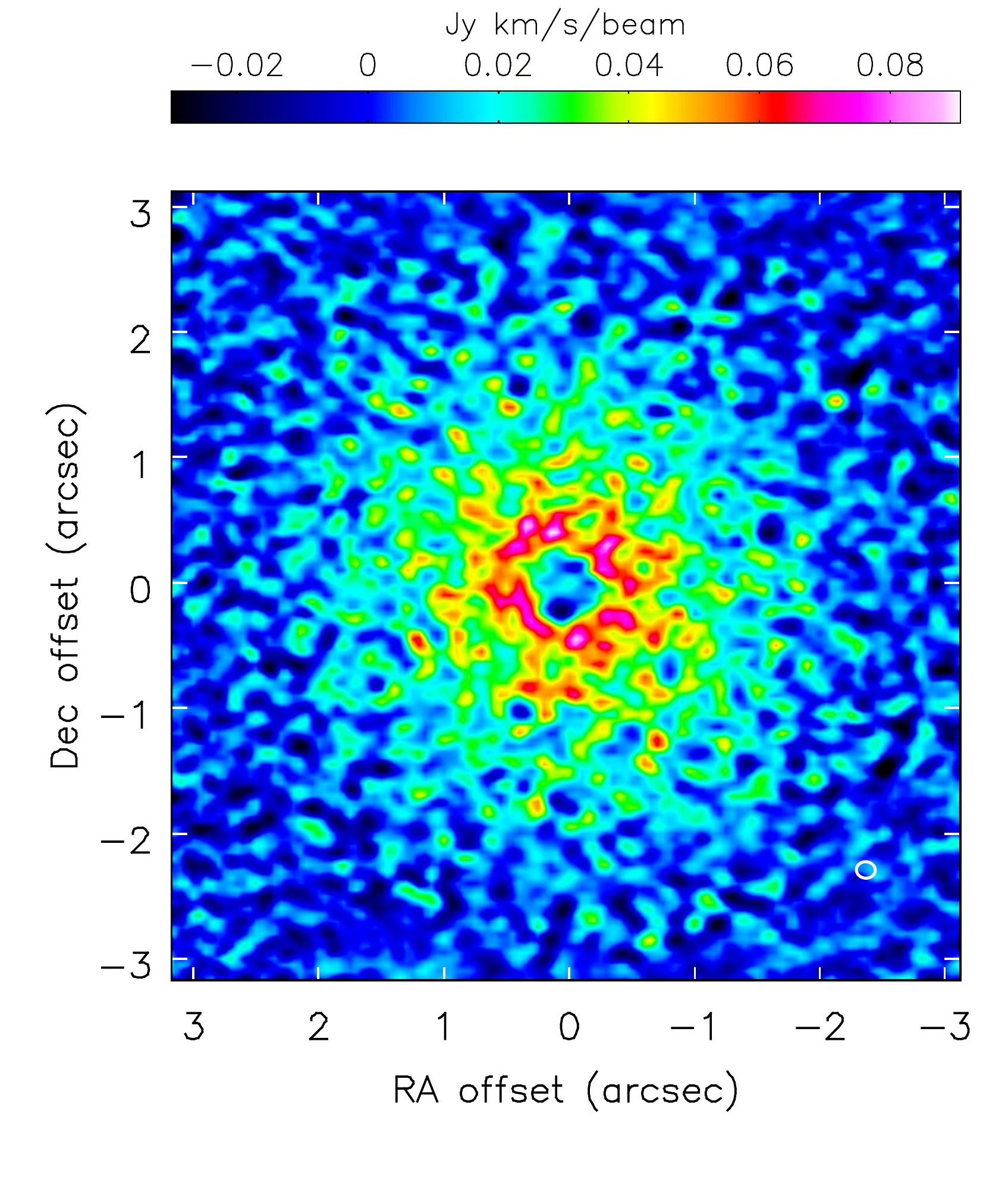}

\caption{{\bf Cavity created by a low-mass planet in a late gas disk.} ALMA synthesised [CI] image (at 492.16 GHz, band 8) of a late disk whose cavity is clearly resolved and carved by a planet at 0.2 arcsec, i.e. 10 au at 50 pc (with 5 hour on source and a beam of 0.12 arcsec).}
\label{figcav}

\end{figure}

With our accretion model, we find that accretion onto planets from late gas disks is very efficient, and for most configurations, a large fraction of the incoming gas is accreted rather than passed on further in (Methods). This means that these gas disks will often be very depleted inwards of a planet and one could infer the presence of a planet from the gas distribution. As the gas extends in the inner region of planetary systems, this new planet detection method could allow us to indirectly detect low-mass planets at a few au or further from their host star. Using the high spatial resolution of ALMA, it would be possible to pinpoint the planet location. In Figure~\ref{figcav}, we show a synthetic ALMA image of the carbon emission of a late gas disk with a terrestrial planet at 10 au from its host star located 50 pc away from Earth (Methods). The cavity is well resolved and the ALMA sensitivity is high enough to detect the signal at the 0.12" resolution. The carbon emission does not seem to extend all the way to the star for the few carbon gas disks\cite{2018ApJ...861...72C,2018arXiv181108439K} observed with ALMA so far but the resolution needs to be pushed further with ALMA to make any strong conclusions.

\begin{addendum}
 \item We thank Giovanni Rosotti, Philippe Thebault and Andrew Shannon for discussions. QK dedicates this paper to Micha\"el.
  \item[Author Contribution] Q.K. led the work, proposed the original idea, wrote the manuscript and produced figures 3 and 4. J.D. coded the model and produced figures 1 and 2. B.C. provided atmospheric parameters to input into the model, expertise in atmosphere observations and produced the synthetic spectra shown in the paper. All authors contributed to the interpretation of the results and commented on the paper.

 \item[Code availability] The particular scripts used for the analysis are made in python and available on reasonable request from the corresponding author.
 \item[Data availability]  The data that support the plots within this paper and other findings of this study are available from the corresponding author upon reasonable request.
 \item[Competing Interests] The authors declare no competing interests.
 \item[Correspondence] Correspondence and requests for materials
should be addressed to Q.K.~(email: quentin.kral@obspm.fr).
\item[Reprints and permissions information] is available at www.nature.com/reprints
\end{addendum}

\begin{methods}

\subsection{Input rate of gas and lifetime of gas disks.}

The input rate of gas from the disk at the planet location is given by $\dot{M}$ in our study. This is the quantity of gas per unit time integrated over the whole disk scaleheight that is transferred radially inwards. These late gas disks are expected to viscously evolve\cite{2013ApJ...762..114X,2016MNRAS.461..845K} (maybe owing to the magnetorotational instability\cite{2016MNRAS.461.1614K}) and there will be a transfer of most of the gas mass inwards (and angular momentum outwards). When steady-state is reached, $\dot{M}$ becomes
equal to the gas mass released in the planetesimal belt, which flows constantly inwards over time and accrete onto a planet or the central star. In this case, we can relate $\dot{M}$ to the surface density of the gas\cite{1974MNRAS.168..603L} $\Sigma$ as $\dot{M}=3 \pi \nu \Sigma$, where $\nu=\alpha c_s H$ is the disk viscosity that can be parametrized with an $\alpha$ value\cite{1973A&A....24..337S} and $c_s$, $H=c_s/\Omega$ are the sound speed and disk scaleheight, respectively, with $\Omega$ the Keplerian frequency. The value of $\alpha$ has been estimated in a few studies by comparing the carbon quantity to what is expected from the measured gas input rate at the planetesimal belt location and it can vary from $10^{-4}$ to 0.1 \cite{2016MNRAS.461..845K,2018arXiv181108439K,2019arXiv190809685M}. Using population synthesis of these gas disks, it seems that all observations so far are consistent with $\alpha$ being close to 0.1 \cite{mari}. An analytic study of the magnetorotational instability in these debris disks\cite{2016MNRAS.461.1614K} show that it could be very active owing to the high ionisation fraction in these disks (compared to protoplanetary disks) and lead to $\alpha$ values close to 0.1, or indeed smaller if non-ideal effects such as ambipolar diffusion are at play.

How fast the disk spreads viscously is set by $\alpha$ such that the viscous timescale $t_{\rm visc}$ equals $r^2 \Omega/(\alpha c_s^2)$, with $r$ the location of the planetesimal belt from which gas is released\cite{1974MNRAS.168..603L}. In Extended data Figure~\ref{figvisc}, we plot $t_\nu$ as a function of $\alpha$ for different radial locations $r$ of the belt and gas temperatures $T$. The viscous timescales are typically $\gtrsim 1$ Myr (even when using the highest $\alpha$ values), meaning that the spreading is often slow but steady state should be reached within the first tens of Myr evolution.

\setcounter{figure}{0}

\begin{figure}
\centering
    \includegraphics[width=11cm]{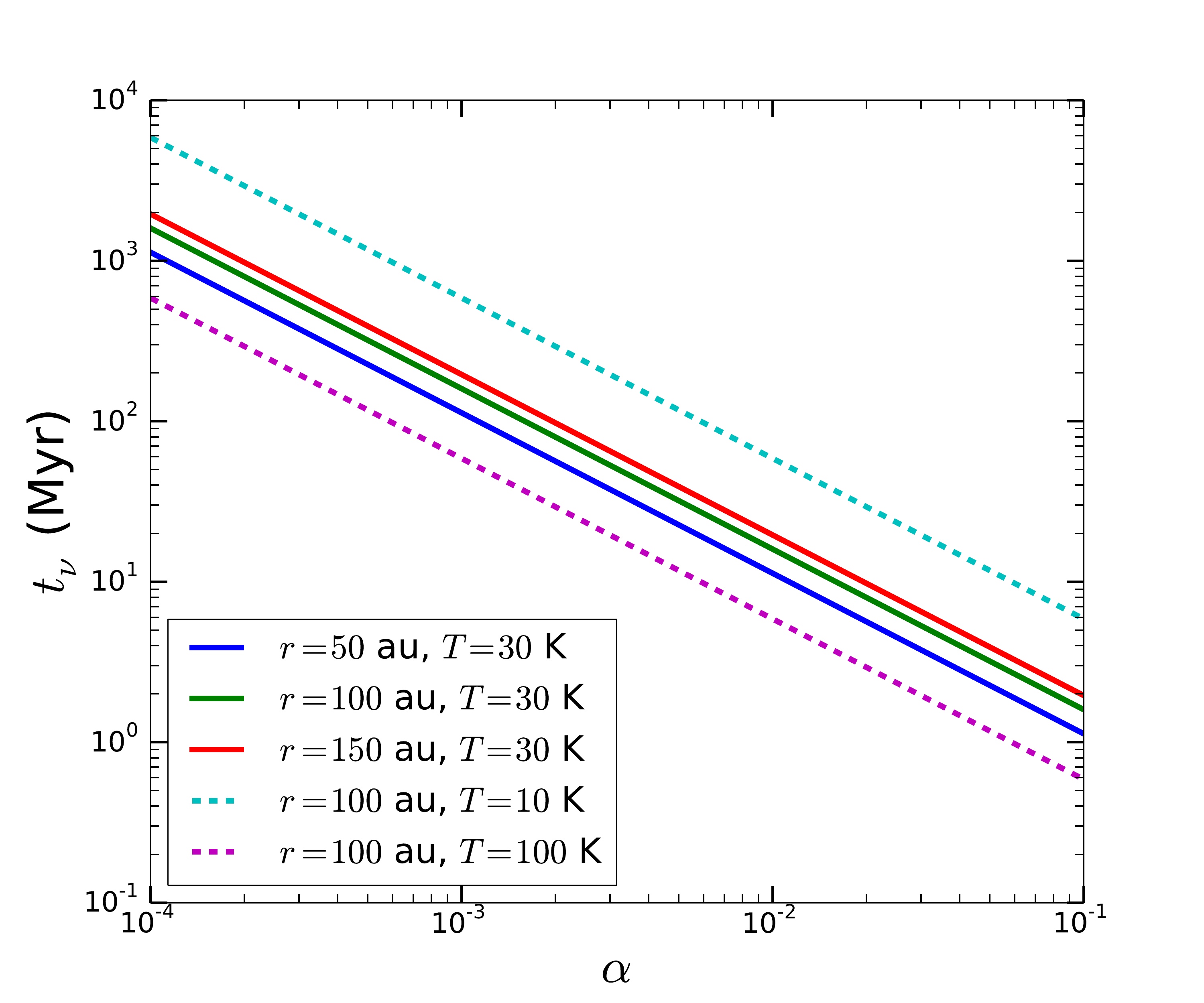}

\caption{Extended data - {\bf Typical viscous evolution timescales.} We plot the viscous timescale $t_\nu$ as a function of the viscous parameter $\alpha$ for different belt locations (50, 100 and 150 au) and different gas temperatures (10, 30 and 100 K).}
\label{figvisc}

\end{figure}

The release rate of gas in late gas disks is expected to decrease with time because planetesimals get destroyed over time and less gas can be released\cite{2007A&A...472..169T,2013A&A...558A.121K,mari}. The $\dot{M}$ we use in this study is an average of the real decreasing input rate over 100 Myr. We suppose that most of the accretion happens when the system is still young (as these belts are mostly observed in systems younger than 100 Myr\cite{2017ApJ...849..123M}) before its planetesimal belt loses too much mass\cite{2008ARA&A..46..339W}, i.e. we do not follow the evolution beyond 100 Myr in our model but it is possible that accretion still happens after 100 Myr at a lower rate. We can quantify the gas release rate as a function of time assuming that gas is released together with dust along the collisional cascade in Solar System comet proportions\cite{2017MNRAS.469..521K}. It is well known that mass loss rates in collisional cascades decrease with time and a simplified model of the evolution of the mass $M_s$ and mass loss rate $\dot{M_s}$ of solids with time is given by\cite{2003ApJ...598..626D,2007ApJ...663..365W}

\begin{align}
M_s(t) &=  \frac{M_{\rm init}}{1+t/t_{\rm col}} \\ 
\dot{M_s}(t) &=  \frac{M_s(t)^2}{M_{\rm init} t_{\rm col}}\\
t_{\rm col}&=\frac{1.4 \times 10^{-9} r^{13/3} \frac{{\rm d}r}{r} D_c Q_D^{\star 5/6}}{ e^{5/3} M_\star^{4/3} M_{\rm tot}}
\end{align}

\noindent with $t_{\rm col}$ the initial collisional timescale of largest planetesimals, where we took typical values\cite{2008ARA&A..46..339W} of largest planetesimal size $D_c=10$ km, belt width d$r=0.5r$, mean eccentricity $e=0.1$, planetesimal strength $Q_D^\star=330$ J/kg and stellar mass $M_\star$=1 M$_\odot$ to compute Extended data Figure~\ref{figmdottime}. We let the disk mean radius $r$ and the initial mass of solids $M_{init}$ be free parameters. We choose to vary $r$ from 50 to 100 au, which is typical of extra-solar Kuiper belt distances\cite{2018ApJ...859...72M} and the initial total mass of solids $M_{\rm init}$ between a Kuiper-belt like mass of 0.1 to 100 M$_\oplus$, which is the maximum mass of solids available in protoplanetary progenitors to debris disks, inferred from sub-mm surveys\cite{2011ARA&A..49...67W}.
We then plot the temporal evolution of the gas release rate $\dot{M}$ in Extended data Figure~\ref{figmdottime} assuming that the gas release rate is 10\% of the dust release rate\cite{2017MNRAS.469..521K}. We can see that $\dot{M}$ remains approximately constant over 100 Myr of evolution except for the most massive belts ($M_{\rm init}$=100 M$_\oplus$) that are closer-in (50 au) and have a much faster collisional evolution. Our approximation of taking $\dot{M}$ constant over 100 Myr works in most cases. However, for the most massive belts, our value of $\dot{M}$ should be understood as being the mean gas input rate integrated over time and one can trace back the $\dot{M}$ value at a given time $t$ using Extended data Figure~\ref{figmdottime} or the set of 3 equations above.

\begin{figure}
\centering
    \includegraphics[width=11cm]{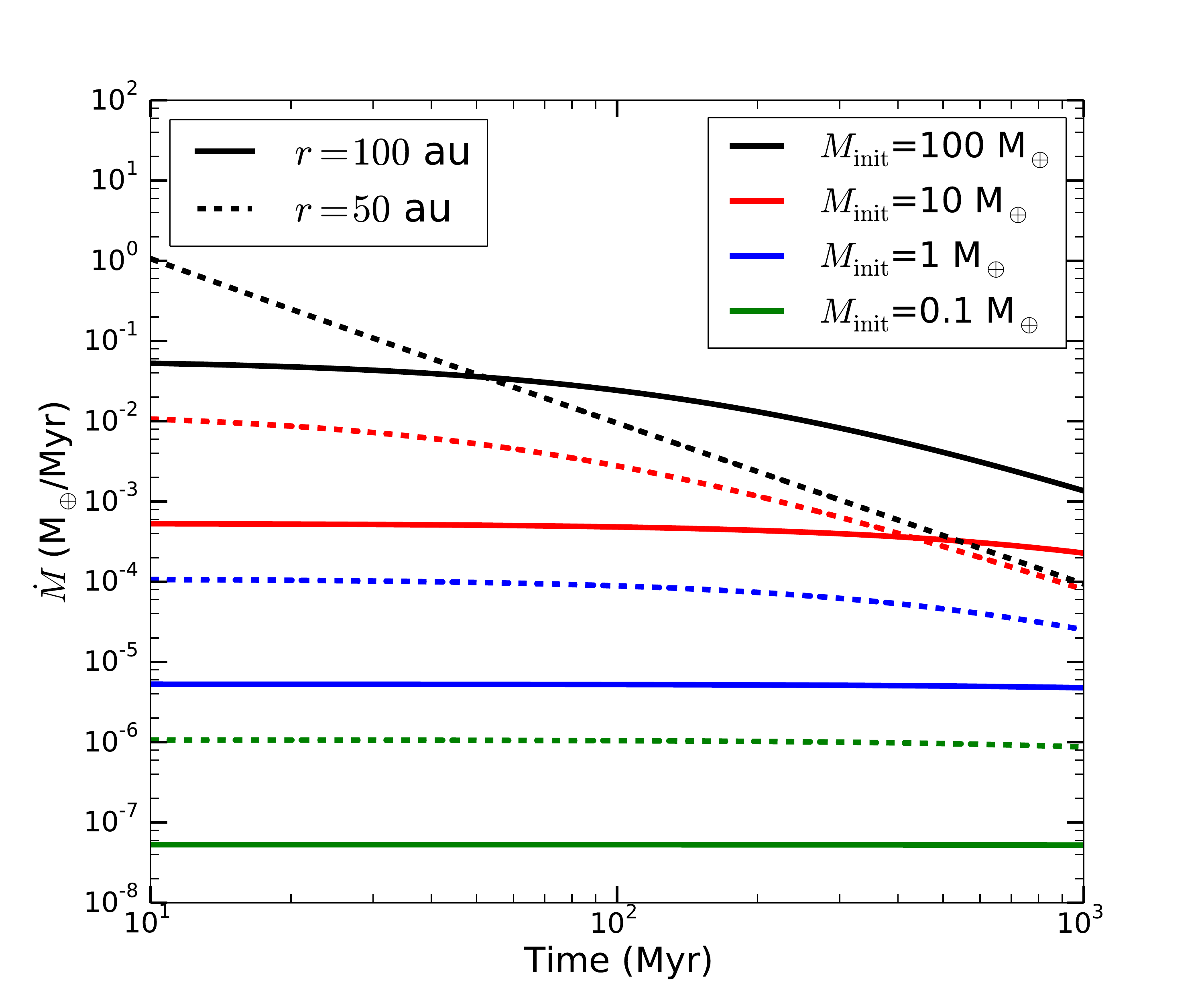}

\caption{Extended data - {\bf Evolution of $\dot{M}$ with time.} We plot the temporal evolution of $\dot{M}$ for different belt locations (50 and 100 au) and different initial belt masses (0.1, 1, 10, and 100 M$_\oplus$).}
\label{figmdottime}

\end{figure}

\subsection{Model.}
The accretion model we used was first developed to explain the formation of Super-Earths in protoplanetary disks\cite{2014ApJ...797...95L}. A good understanding of the way accretion works was obtained through analytical model fitting to numerical simulations\cite{2015ApJ...811...41L}. It is shown that the accretion rate hardly depends on the gas density from which it accretes but rather on the planet's ability to cool (or radiate away its energy). The more a planet can cool, the more it contracts, emptying parts of its Hill sphere, which get refilled very rapidly and the same process can happen again on a Kelvin-Helmholtz time. We also note that this accretion model is valid even in low-gas density environments where planetary atmospheres are optically thin to incident starlight\cite{2018MNRAS.476.2199L}. Therefore, a good estimate for the accretion timescale is given by the cooling timescale equal to

\begin{equation}
t_{\rm cool}=|E|/L_{\rm cool},
\end{equation}

\noindent where $E$ is the atmosphere energy and $L_{\rm cool}$ its luminosity\cite{2015ApJ...811...41L}. From that timescale, we calculate the gas-to-core ratio GCR=$M_{\rm gas}/M_{\rm core}$ as a function of time $t$

\begin{equation}
\label{eqGCR}
{\rm GCR}(t)=\frac{-bt+\sqrt{b^2t^2-4abt}}{2a},
\end{equation}

\noindent where $a=3 f_E k_B^{1+\frac{1}{\gamma -1}}(4\pi)^{\frac{1}{3}\left(3-\frac{1}{\gamma - 1}\right)}\rho_b^{\frac{1}{3}\left(3-\frac{1}{\gamma -1}\right)}\kappa_{\rm rcb}$, and \\ $b=-4^2\pi ^2 G^{1+1/(\gamma-1)}\sigma T_{\rm rcb}^{3-1/(\gamma -1)}(\mu_{\rm rcb}m_{\rm H})^{1+1/(\gamma -1)}\nabla_{\rm ad}^{1+1/(\gamma -1)}M_{\rm core}^{\frac{2}{3}\left(\frac{1}{\gamma -1}-1\right)}3^{\frac{1}{3}\left(3-\frac{1}{\gamma -1}\right)}$. The term $f_E$ equals $G(4\pi \rho_b / 3)^{1/3}$, and $\rho_b$ is the bulk density of the core, $M_{\rm core}$ is the core mass, $\nabla_{\rm ad}=(\gamma-1)/\gamma$ is the adiabatic gradient, $T_{\rm rcb}$, $\mu_{\rm rcb}$, $\kappa_{\rm rcb}$ are the temperature, mean molecular weight and atmospheric opacity at the radiative/convective boundary (rcb). Finally, $m_{\rm H}$ is the proton mass, $k_B$ the Boltzmann constant, $G$ the gravitational constant and $\sigma$ the Stefan-Boltzmann constant. This new analytic expression given by Eq.~\ref{eqGCR} is otherwise and do not make the assumption that GCR $\ll$ 1 as in previous work\cite{2015ApJ...811...41L}. Unless otherwise stated, we fix all values as given in the reference paper\cite{2015ApJ...811...41L}. To be realistic, we use real opacities derived for highly metal enriched (100 times solar abundances) Super-Earths\cite{2014ApJS..214...25F} and in this case GCR$(t)$ must be obtained numerically (because now opacities vary with GCR). We fix $\gamma=1.4$ and the mean molecular weight as being 28, which corresponds to an atmosphere accreting a majority of CO as would be the case in our scenario\cite{2018arXiv181108439K}. We note that in non-shielded disks (i.e. disks not massive enough for carbon production to be sufficiently high that neutral carbon shields CO from photodissociating\cite{2018arXiv181108439K}), $\mu$ is closer to 14, i.e., dominated by carbon and oxygen\cite{2016MNRAS.461..845K,2017ApJ...842....9M} rather than CO but our final results are not so much affected by the value of $\mu$ (except for the very high $\dot{M}$ values) as demonstrated further in Extended data Figure~\ref{figmuc}. There could also be some other molecules released in small quantities in the belt, such as CN, but only CO is detected so far.

We find that GCR$(t)$ can be simplified in the limit where $-a/(bt)  \gg 1$ or $-a/(bt)  \ll 1$. For $-a/(bt)  \gg 1$, we find GCR $=\sqrt{-bt/a}$ and when $-a/(bt)  \ll 1$, GCR $=-bt/a$. The latter regime has never been studied so far as it only appears at large $t$ and large $\mu$, which is typical of late gas disks but not of protoplanetary disks.

\begin{figure*}
\centering
    \includegraphics[width=8cm]{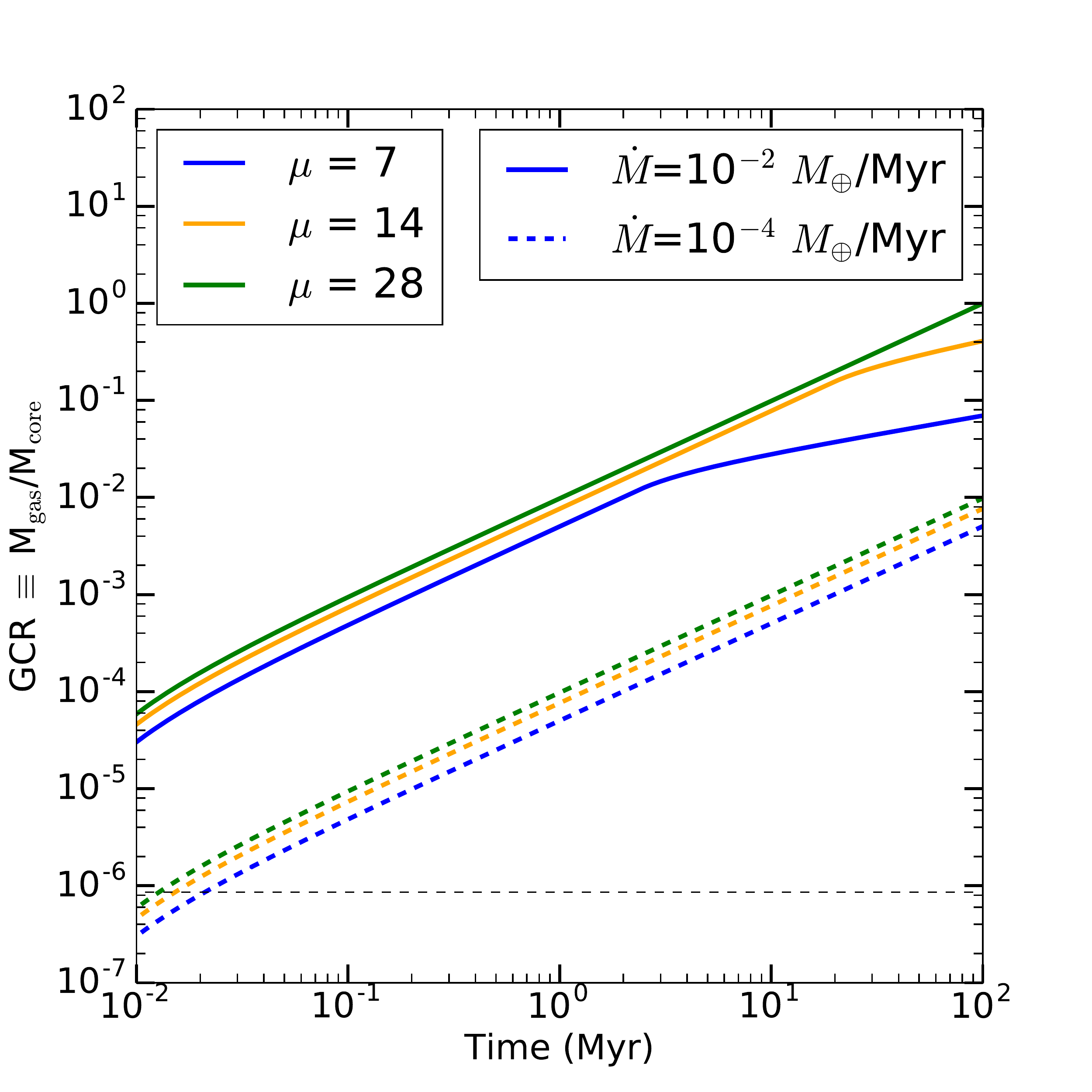}

\caption{Extended data - {\bf GCR for different values of mean molecular weight $\mu$.} Temporal evolution of the gas-to-core ratio (GCR) with varying $\mu$. We note that the GCR grows more slowly than expected for the cases $\mu=7$ and 14 when $\dot{M}=10^{-2}$ M$_\oplus$/Myr, which is because the theoretical cooling accretion rate becomes smaller than $10^{-2}$ M$_\oplus$/Myr for lower values of $\mu$ (see Fig.~\ref{figmdotth}). We also see that for lower values of $\mu$, the accretion is less efficient from the start because the gas disk scaleheight is higher and less gas is accreted (see Fig.~\ref{figrh}).}
\label{figmuc}
\end{figure*}

The maximum rate at which a planet can accrete an atmosphere is given by $\dot{M}_{\rm gas}$, which we obtain by deriving $M_{\rm gas}$=GCR$(t) M_{\rm core}$. We find

\begin{equation}
\label{eqmdotgas}
\dot{M}_{\rm gas}=\frac{M_{core}}{2a}\times \left[-b+\frac{1}{2}\times\left(b^2t^2-4abt\right)^{-1/2}\times\left(2b^2t-4ab\right)\right],
\end{equation}

\noindent where $a$ and $b$ are defined above. In the previous Eq.~\ref{eqmdotgas}, we have assumed a constant opacity but in reality the opacity varies with time as the GCR increases. In our code, we compute this derivative ($\dot{M}_{\rm gas}$) numerically to take this complexity into account. We also derive $\dot{M}_{\rm gas}$ analytically for the case where $\kappa$ depends on time. The opacity can be parametrized\cite{2014ApJS..214...25F} as $\kappa \propto \kappa_0 \rho_{\rm rcb}^\alpha T_{\rm rcb}^\beta Z^\delta$, where the time dependence comes in the density at the rcb $\rho_{\rm rcb}$, which is proportional to GCR\cite{2015ApJ...811...41L}. We find that there are also two regimes depending on the value of $-a/(bt)$. If $-a/(bt)  \gg 1$ then $\dot{M}_{\rm gas}=M_{\rm core}/(2+3\alpha) \sqrt{-b/(a t)}$ and when $-a/(bt)  \ll 1$ then $\dot{M}_{\rm gas}=-M_{\rm core}/(1+\alpha) (b/a)$. 
For the case where $-a/(bt)  \gg 1$, we thus find

\begin{equation}
\label{eqmdotgasscale0a}
\begin{split}
M_{\rm gas} \propto & \, \left[ T_{\rm rcb}^{3-\beta-(1+\alpha)/(\gamma -1)} (\mu_{\rm rcb} \nabla_{\rm ad})^{1+(1+\alpha)/(\gamma-1)} \rho_b^{-4/3-\alpha+1/3(1+\alpha)/(\gamma-1)}  M_{\rm core}^{2/3(1+\alpha)/(\gamma-1)+4/3+2\alpha}  \right. \\ \left. Z^{-\delta} \kappa_0^{-1} t \right]^{1/(2(1+\alpha))} \\
\dot{M}_{\rm gas} \propto & \, \left[ T_{\rm rcb}^{3-\beta-(1+\alpha)/(\gamma -1)} (\mu_{\rm rcb} \nabla_{\rm ad})^{1+(1+\alpha)/(\gamma-1)} \rho_b^{-4/3-\alpha+1/3(1+\alpha)/(\gamma-1)} M_{\rm core}^{2/3(1+\alpha)/(\gamma-1)+4/3+2\alpha} \right. \\ \left. Z^{-\delta} \kappa_0^{-1} t^{-1-2\alpha} \right]^{1/(2(1+\alpha))} 
\end{split}
\end{equation}

\noindent and for the case $-a/(bt)  \ll 1$, we find

\begin{equation}
\label{eqmdotgasscale0b}
\begin{split}
M_{\rm gas} \propto & \, \left[ T_{\rm rcb}^{3-\beta-(1+\alpha)/(\gamma -1)} (\mu_{\rm rcb} \nabla_{\rm ad})^{1+(1+\alpha)/(\gamma-1)} \rho_b^{-4/3-\alpha+1/3(1+\alpha)/(\gamma-1)}  M_{\rm core}^{2/3(1+\alpha)/(\gamma-1)+1/3+\alpha} \right. \\ \left. Z^{-\delta} \kappa_0^{-1} t \right]^{1/(1+\alpha)} \\
\dot{M}_{\rm gas} \propto & \, \left[ T_{\rm rcb}^{3-\beta-(1+\alpha)/(\gamma -1)} (\mu_{\rm rcb} \nabla_{\rm ad})^{1+(1+\alpha)/(\gamma-1)} \rho_b^{-4/3-\alpha+1/3(1+\alpha)/(\gamma-1)} M_{\rm core}^{2/3(1+\alpha)/(\gamma-1)+1/3+\alpha} \right. \\ \left. Z^{-\delta} \kappa_0^{-1} t^{-\alpha} \right]^{1/(1+\alpha)}\end{split}
\end{equation}

\noindent Taking an opacity with $\alpha=0.6$, $\beta=2.2$ and $\delta=1$ valid for dust free atmospheres beyond 1 au\cite{2015ApJ...811...41L} and that $\gamma=1.4$ (for CO atmospheres), we find for the case $-a/(bt)  \gg 1$ that 

\begin{equation}
\label{eqmdotgasscale1}
\begin{split}
M_{\rm gas} \propto & \, T_{\rm rcb}^{-1} \,\mu_{\rm rcb}^{1.6} \,\rho_b^{-0.19} \,M_{\rm core}^{1.6} \,Z^{-0.3} \,\kappa_0^{-0.6} \,t^{0.3}\\
\dot{M}_{\rm gas} \propto & \, T_{\rm rcb}^{-1} \,\mu_{\rm rcb}^{1.6} \,\rho_b^{-0.19} \,M_{\rm core}^{1.6} \,Z^{-0.3} \,\kappa_0^{-0.6} \,t^{-0.7}
\end{split}
\end{equation}

\noindent and for the case $-a/(bt)  \ll 1$, we find

\begin{equation}
\label{eqmdotgasscale2}
\begin{split}
M_{\rm gas} \propto & \, T_{\rm rcb}^{-2} \,\mu_{\rm rcb}^{3.1} \,\rho_b^{-0.38} \,M_{\rm core}^{2.3} \,Z^{-0.6} \,\kappa_0^{-0.6} \,t^{0.6}\\
\dot{M}_{\rm gas} \propto & \, T_{\rm rcb}^{-2} \,\mu_{\rm rcb}^{3.1} \,\rho_b^{-0.38} \,M_{\rm core}^{2.3} \,Z^{-0.6} \,\kappa_0^{-0.6} \,t^{-0.4}
\end{split}
\end{equation}

In late gas disks, the total gas mass is much smaller than in protoplanetary disks and it may happen that the gas available per unit time is smaller than $\dot{M}_{\rm gas}$. Therefore, in our code, at each time-step, we compare the gas crossing rate $\dot{M}$ to $\dot{M}_{\rm gas}$. 
For most cases studied in this paper (except the Mars-mass planet case, see Fig.~\ref{figmars}), we find that $\dot{M}$ is indeed lower than $\dot{M}_{\rm gas}$ and the accretion is limited by the quantity available rather than by the planet cooling. For the cases where $\dot{M} < \dot{M}_{\rm gas}$, the mass that accumulates is lower than the theoretical mass $M_{\rm gas}$ given by Eq.~\ref{eqGCR} so that when computing $\dot{M}_{\rm gas}$ one should take the theoretical accretion rate for the mass that actually accumulated rather than the theoretical mass, which we do in our code. For the regime where $-a/(bt)  \gg 1$ this means that at a given time $t$, one should take the accretion rate at time $t'=t(\dot{M} t /M_{\rm gas})^{2+\alpha}$, which gives an accretion rate which is higher by a factor $M_{\rm gas}/(\dot{M} t)$. We show the theoretical $\dot{M}_{\rm gas}$ in Extended data figure~\ref{figmdotth} where we plot the temporal evolution of $\dot{M}_{\rm gas}$ (taking into account that the opacity varies with time) for different values of planet semi-major axes $a_{\rm pl}$, atmospheric mean molecular weight $\mu$ and core mass $M_{\rm core}$, corresponding to the typical range of values that we are interested in in our study.

\begin{figure}
\centering
    \includegraphics[width=7cm]{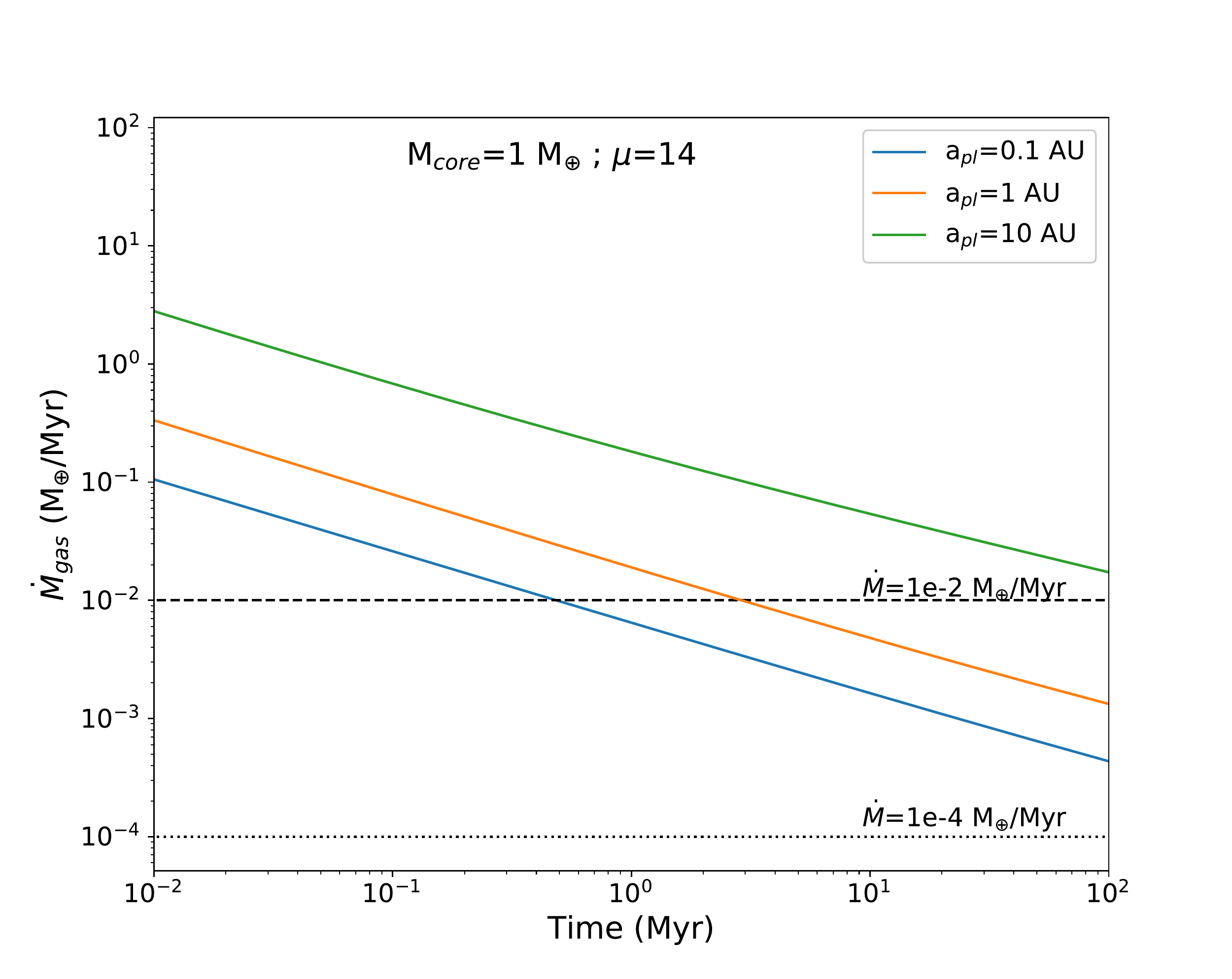}
    \includegraphics[width=7cm]{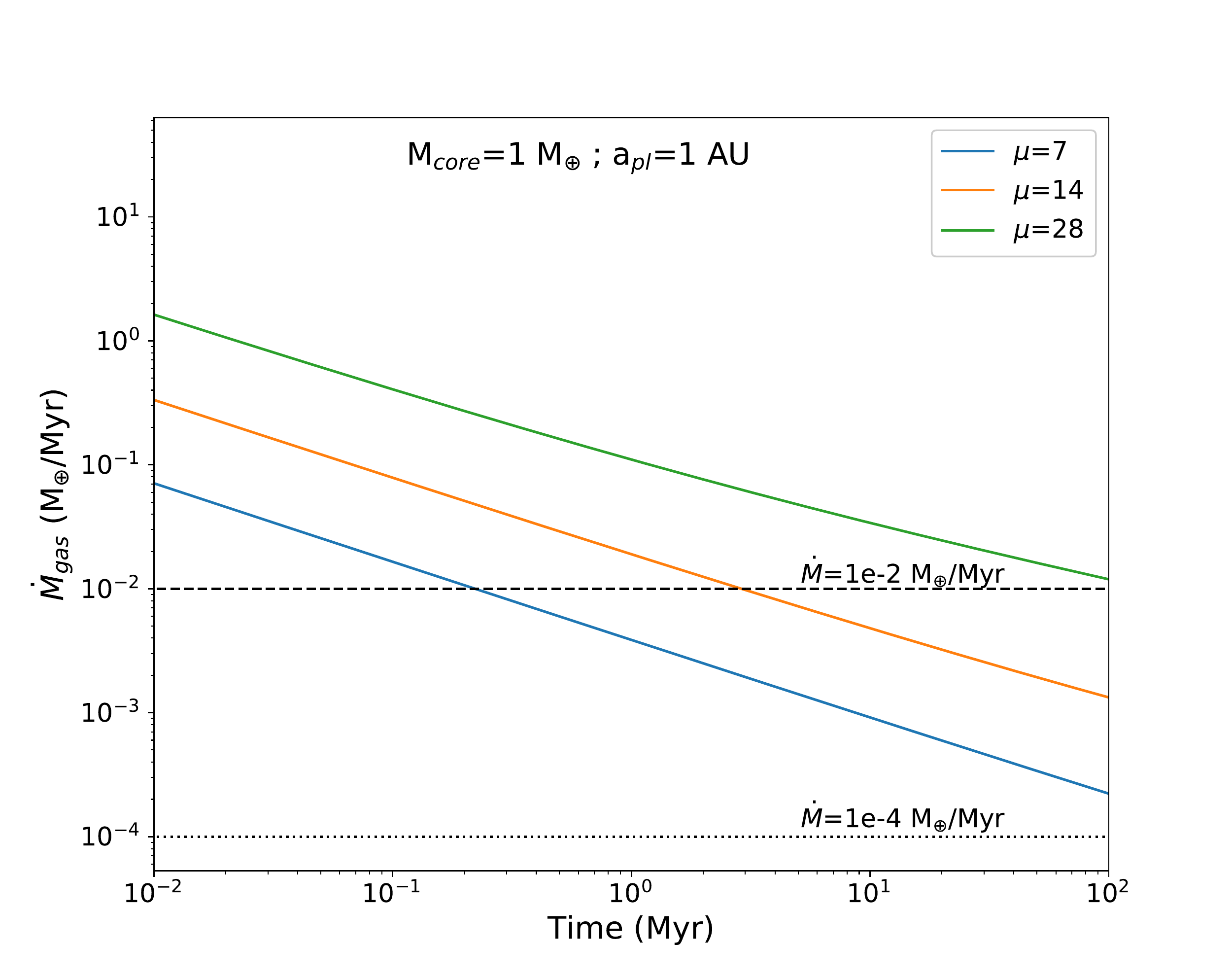}
    \includegraphics[width=7cm]{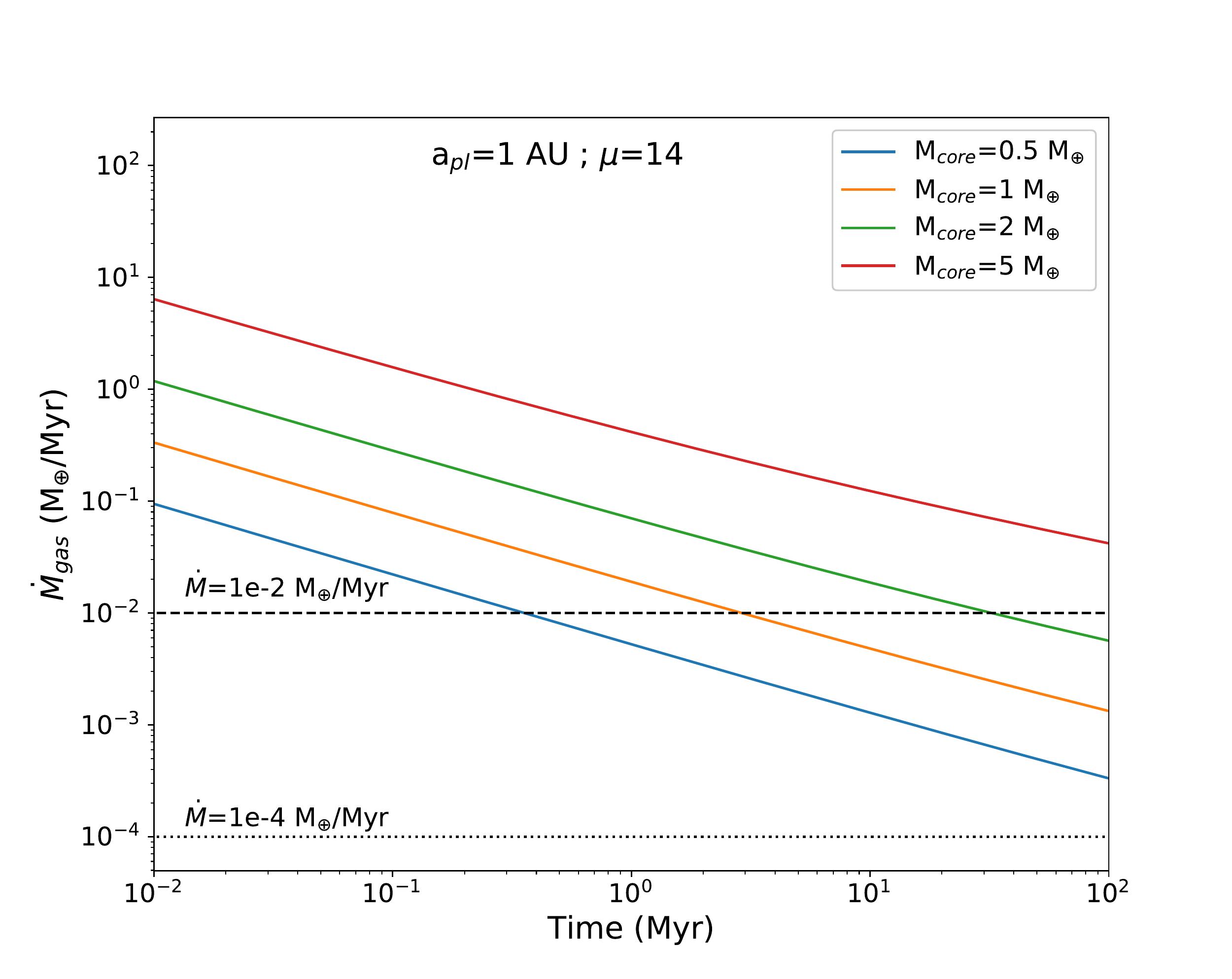}

\caption{Extended data - {\bf Potential accretion rate on a planet Vs. available accretion rates.} We plot the temporal evolution of the potential theoretical accretion rate on a planet $\dot{M}_{\rm gas}$ (numerical derivative with $\kappa$ varying with time) for different values of planet semi-major axes $a_{\rm pl}$, atmospheric mean molecular weight $\mu$ and core mass $M_{\rm core}$. The fiducial model is $a_{\rm pl}=1$ au, $\mu=14$, and $M_{\rm core}=1$ M$_\oplus$. We overplot horizontal lines with different input rate values of our parameter $\dot{M}$, including the case of $10^{-2}$ M$_\oplus$/Myr over 100 Myr to verify whether in the cases studied in this paper the theoretical accretion rate is higher than our $\dot{M}$ parameter (which is always the case for $\mu=28$, large $M_{\rm core}$ or distant planets). We note that for the $\mu=28$ case, the green lines become less steep at large $t$. This is because as $t$ increases, one reaches the second regime for which $-a/(bt)  \ll 1$, where $\dot{M}_{\rm gas}$ scales as $t^{-0.4}$ instead of $t^{-0.7}$ in the other regime (see Eqs~\ref{eqmdotgasscale1} and \ref{eqmdotgasscale2}). We also note that for the case at 0.1 au (for which $T=1000$ K), the opacity varies more slowly with $T$ for high enough densities and $\beta$ becomes smaller\cite{2015ApJ...811...41L}, hence leading to a higher $\dot{M}_{\rm gas}$.}
\label{figmdotth}

\end{figure}

We also compare the Hill radius $R_H$ (equal to $a_{\rm pl} [M_{\rm pl}/(3 M_\star)]^{1/3}$) of the planet (of semi-major axis $a_{\rm pl}$ and mass $M_{\rm pl}$) with the scale height $H$ of the disk. For very small Earth-like planets (see Extended data figure~\ref{figrh}), it can happen that $R_H<H$ and in this case, some gas crossing at a rate $\dot{M}$ cannot be accreted by the planet. We then recalculate a new $\dot{M}$ that can be accreted by only considering the $\dot{M}$ that crosses the planet's Hill sphere rather than the whole scaleheight. To calculate the quantity of gas that cannot be accreted, we assume a sphere of radius $H$ on top of the sphere of radius $R_H$, and take out the parts of the large $H$ sphere with a height greater than $R_H$. This gives a new $\dot{M}_{\rm new}$ value that is $3/2 (R_H/H)-1/2(R_H/H)^3$ of the full $\dot{M}$. Note that for cases that reach GCR values approaching 1, $R_H$ can become larger and $R_H/H$ becomes greater, hence enhancing accretion. Therefore, we find that $\dot{M}_{\rm new}$ scales roughly with $R_H/H$ rather than its square as would be expected from Bondi-accretion or dividing the volume of two tori of radii $R_H$ and $H$, respectively. This is because the orbital timescale of the planet is much faster than the viscous drift timescale of the gas flow and the new $\dot{M}$ will be approximately equal to the ratio of cylindrical collisional cross section of the planet $2 \pi a_{\rm pl} R_H$ and the cylindrical flow cross section $2 \pi a_{\rm pl} H$, which equals $R_H / H$.

\begin{figure}
\centering
    \includegraphics[width=11cm]{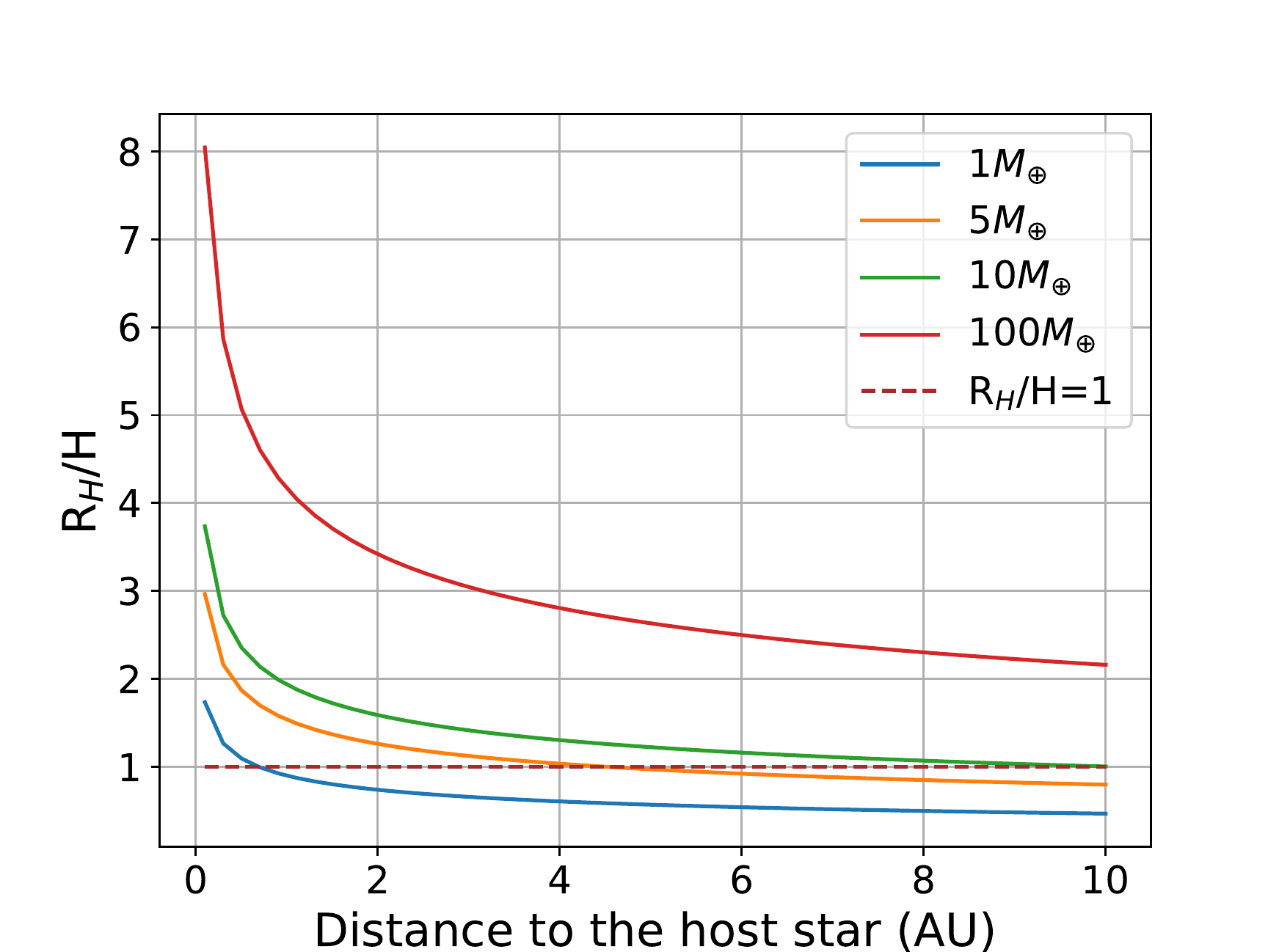}

\caption{Extended data - {\bf Is a planet accreting all gas flowing through the disk?} We plot the Hill sphere radius-to-scaleheight ratio Vs. the distance to host star. For the lowest-mass planets, the Hill sphere radius can be smaller than the disk scaleheight and gas can flow inwards rather than being accreted.}
\label{figrh}

\end{figure}

Furthermore, we compare our typical accretion rate $\dot{M}$ with a Bondi-like accretion that may be relevant in the regime where $R_H<H$. Bondi-like accretion happens at a rate $\dot{M}_B=4 \pi R_H^2 \Sigma \Omega$, where $\Sigma$ is the local secondary disk gas surface density (with $\Sigma=\dot{M}/(3 \pi \nu)$ at steady state), and $\Omega$ is the Keplerian frequency. We find that  $\dot{M}_B/\dot{M}=4/(3\alpha) (R_H/H)^2$ or $\dot{M}_B/\dot{M}_{\rm new} \sim (1/\alpha) (R_H/H)$, with $\alpha$ typically between $10^{-4}$ and 0.1 as explained earlier\cite{2016MNRAS.461..845K,2016MNRAS.461.1614K}. It means that in most cases as $R_H/H>0.1$, the Bondi accretion rate is higher than the rate at which gas is delivered, and one is still limited by $\dot{M}$. For instance, for a 1 M$_\oplus$ planet at 1 au (with $R_H/H\sim0.9$ as shown in Extended data Figure~\ref{figrh}), we obtain a new input rate that is $\sim$0.98 times of the full $\dot{M}$. This value is therefore usually very close to 1 and does not make important differences even when taking extreme cases such as a Mars-mass (0.1 M$_\oplus$) planet at Mars distance (1.5 au) or a 0.5 M$_\oplus$ planet at 10 au, for which we find $R_H/H\sim0.37$ for both cases, and the new input rate $\dot{M}_{\rm new}$ is 0.53 of the full $\dot{M}$, therefore lowering the final masses accreted by a factor 2 on such planets. We show the results for these two planets in Extended data Figure~\ref{figmars}. We end up with GCR values very similar to the case in Figure~\ref{figgcr} (1 M$_\oplus$) for the case 0.5 M$_\oplus$ because although the final gas mass is divided by 2, the core mass is also twice smaller, hence the GCR is similar. Owing to the much lower core mass of the Mars-mass case, the final GCR is 5 times higher than in the 1 M$_\oplus$ case except for the highest input rates where the theoretical cooling accretion rate becomes smaller than $\dot{M}$ and the accreted mass becomes smaller, explaining the shallower slope of GCR at large $t$. For cases with $R_H/H>1$, e.g. planets with more massive cores, the results are similar to the case shown in Figure~\ref{figgcr} after a correction of the GCR to take account of the new $M_{\rm core}$ value (i.e. the GCR goes down by a factor $\propto M_{\rm core}$) but we note that in this case the hydrodynamics of the flow can be complicated owing to shock waves developing near the planet-disk interface\cite{2012ApJ...747...47T}.

\begin{figure}
\centering
    \includegraphics[width=8cm]{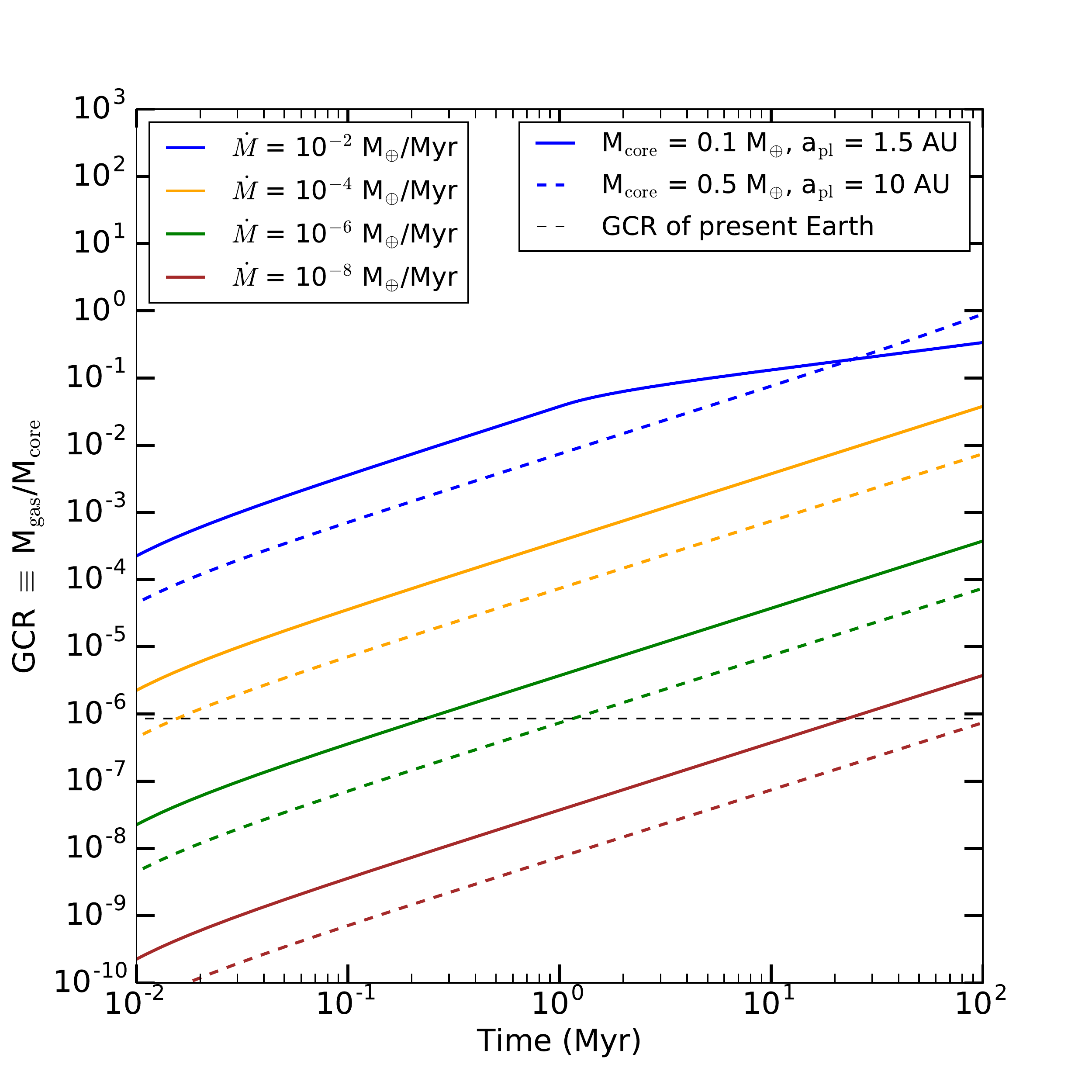}
    \includegraphics[width=8cm]{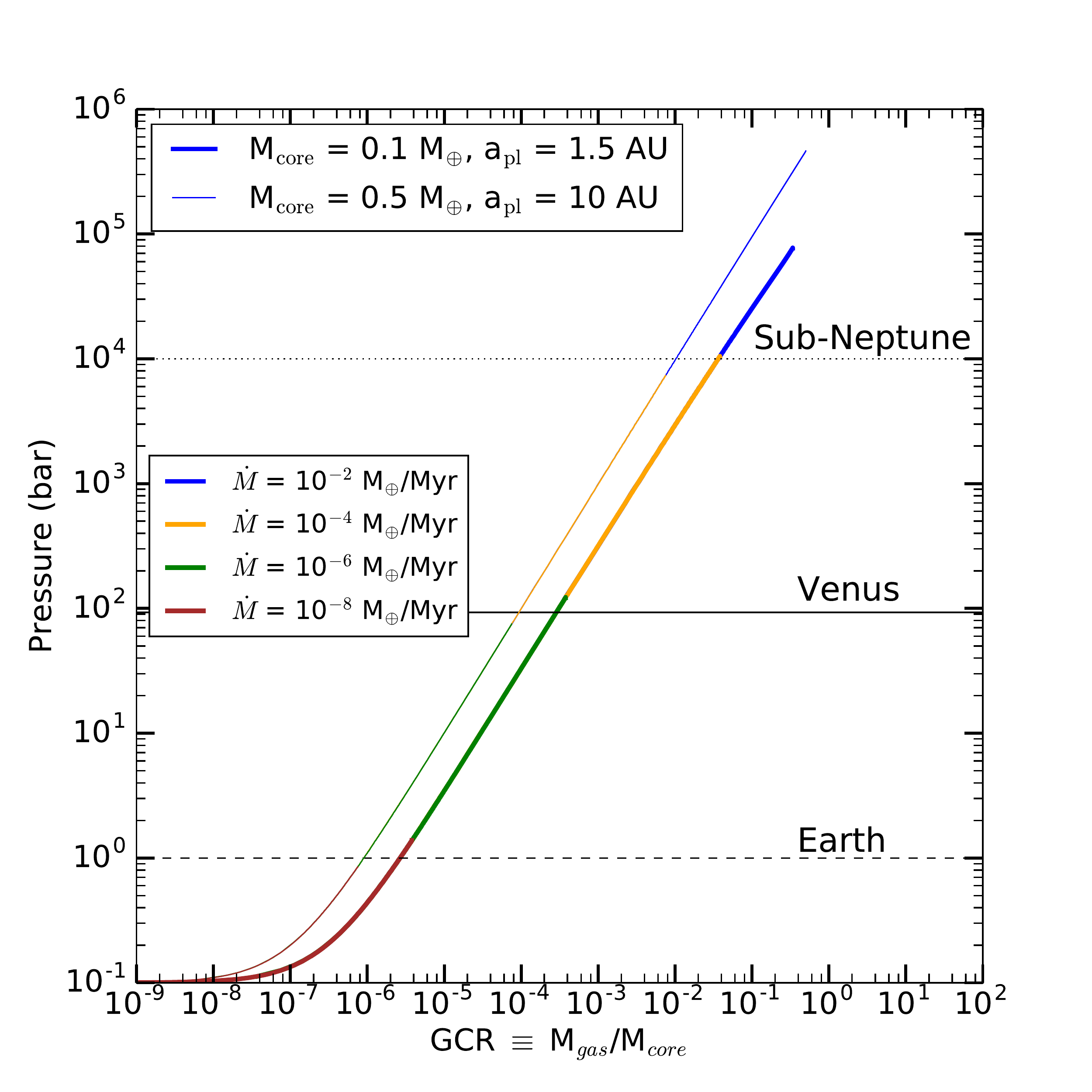}

\caption{Extended data - {\bf GCR and pressures for a Mars-mass and a distant planet.} Temporal evolution of the gas-to-core ratio (left) and pressure (right) for a Mars-mass (0.1$\,$M$_\oplus$ at 1.5 au) planet and a distant planet (10 au) with a core mass of $0.5 \, $M$_\oplus$ up to a GCR of 0.5. In the pressure plot (right), the thick solid line is for the Mars-like planet case and the thinner line is for the core of mass $0.5$ M$_\oplus$ at 10 au. We note that for the Mars-mass case, GCR grows more slowly than expected when $\dot{M}=10^{-2}$ M$_\oplus$/Myr, which is because the theoretical cooling accretion rate becomes smaller than $10^{-2}$ M$_\oplus$/Myr in this case (see Fig.~\ref{figmdotth}).}
\label{figmars}
\end{figure}

To verify that the Hill radius is the radius of interest, which sets the length scale of accretion as assumed here, we also compute the Bondi radius (equal to $2 G M_{\rm pl}/c_s^2$) for a large variety of planet masses and planet semi-major axes. Extended data Figure~\ref{fighill} shows that indeed the Bondi radius is always greater than the Hill radius in our study, confirming that the Hill radius should be used in previous computations.

\begin{figure}
\centering
    \includegraphics[width=11cm]{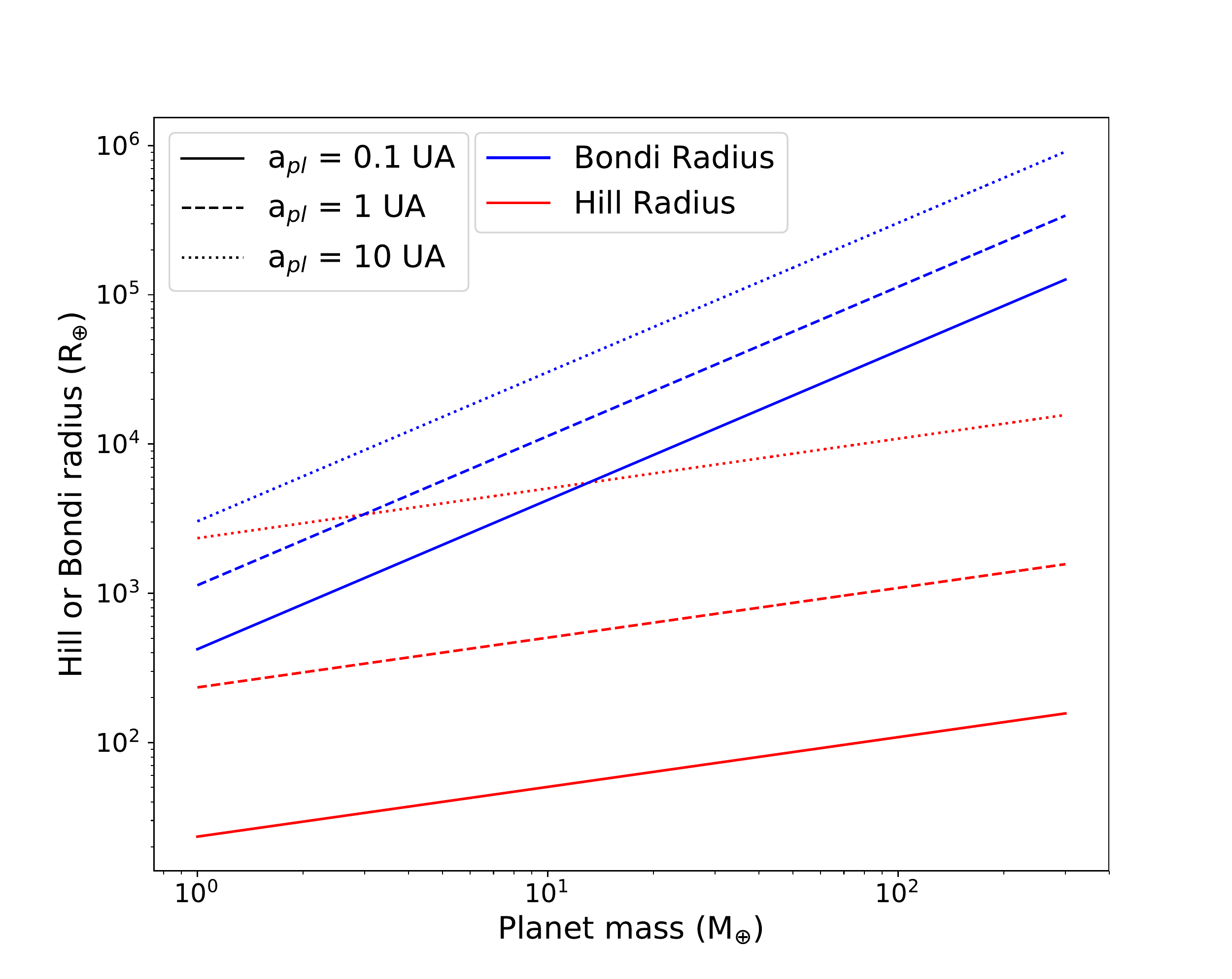}

\caption{Extended data - {\bf Hill Vs. Bondi radii.} Hill and Bondi radii Vs. planet mass for different planet semi-major axes.}
\label{fighill}

\end{figure}

To compute the planet's bulk density given its GCR and core mass, we improve upon previous work\cite{2018ApJ...865...20D} and do not assume that the gravity $g$ is constant with pressure so that the thickness of the adiabatic convective region of the atmosphere is given by (when $R_{\rm core}>c_p/g_{\rm batm}(T_{\rm batm}-T_{\rm atm}$))

\begin{equation}
d_{\rm atm}=\frac{c_p}{g_{\rm batm}} \left( T_{\rm batm}-T_{\rm atm} \right) \frac{1}{1-c_p/(R_{\rm core}g_{\rm batm})(T_{\rm batm}-T_{\rm atm})},
\end{equation}

\noindent where $P_{\rm atm}=0.1$ bar typical of the pressure at which the atmosphere transitions from adiabatic to isothermal\cite{2014NatGe...7...12R}, $T_{\rm batm}=T_{\rm atm} (P_{\rm batm}/P_{\rm atm})^\kappa$, where $\kappa=2/(2+n)$, with $n$ the number of degrees of freedom equal to 5 for diatomic gases, with $T_{\rm atm}=T_\star/2 \sqrt{R_\star/(2a)}$ ($T_\star$ and $R_\star$ are the temperature and radius of the host star), and we calculate $P_{\rm batm}$ by integrating ${\rm d}m=4 \pi (R_{\rm core}+z)^2 {\rm d}P/g(P)$ from $P_{\rm atm}$ up to a value $P$ for which the atmosphere mass is $M_{\rm atm}$=GCR$M_{\rm core}$. In the integral, we use that $g(z)=g_{\rm batm}/(1+z/R_{\rm core})^2$ and $z=T_{\rm atm} c_p/g_{\rm batm} ((P_{\rm batm}/P_{\rm atm})^{\kappa}-(P/P_{\rm atm})^{\kappa})/(1-c_p T_{\rm atm}/(R_{\rm core} g_{\rm batm})((P_{\rm batm}/P_{\rm atm})^{\kappa}-(P/P_{\rm atm})^{\kappa}))$, where $c_p$ is the heat capacity, $T_{atm}$ the atmosphere surface temperature, $g_{batm}$ and $T_{batm}$ the gravity and the temperature at the bottom of the atmosphere, respectively. Since $z$ depends on $P_{\rm batm}$, we calculate $P_{\rm batm}$ by iteration, starting from an initial value equal to $M_{\rm atm}g_{\rm batm}/(4\pi R_{\rm core}^2)$. We then add the isothermal part of the atmospheric thickness from 100 to 20 mbar: $R^\star \, T_{\rm atm}/(g({\rm 100 mbar}) \mu) \ln(100/20)$ (where $R^\star$ is the universal gas constant) assuming that the transit radius is typically observed at around 20 mbar\cite{2018ApJ...865...20D}.

\subsection{Accretion onto a pre-existing Earth-like atmosphere.}

In Extended data Figure~\ref{figpre}, we show what happens when an atmosphere starts growing from a pre-existing Earth-mass atmosphere rather than an empty atmosphere as in Figures~\ref{figgcr} and ~\ref{figdens}. The results are the same, with all the lines being just shifted up, starting at the pre-existing level.

\begin{figure}
\centering
    \includegraphics[width=8cm]{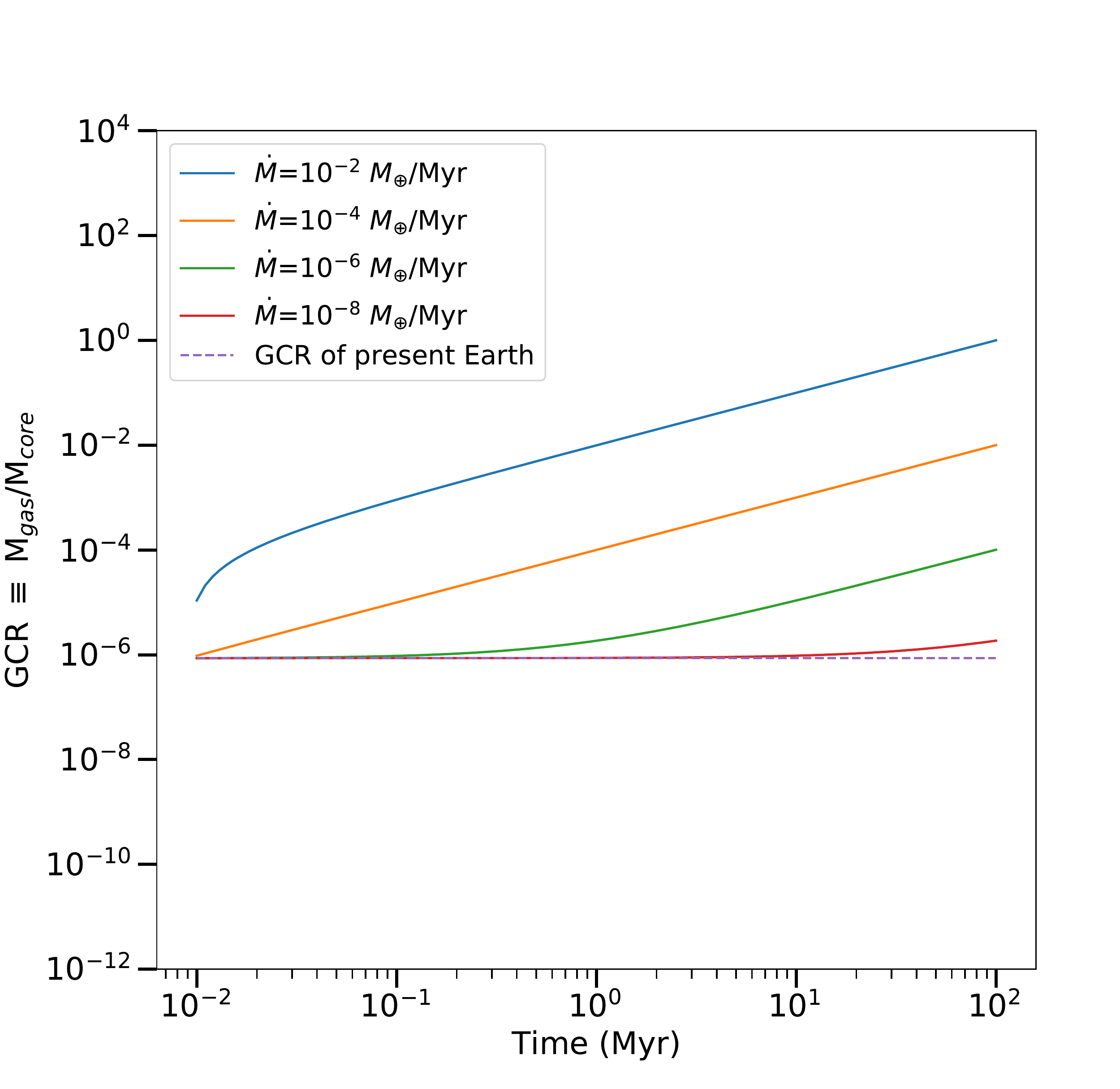}
    \includegraphics[width=8cm]{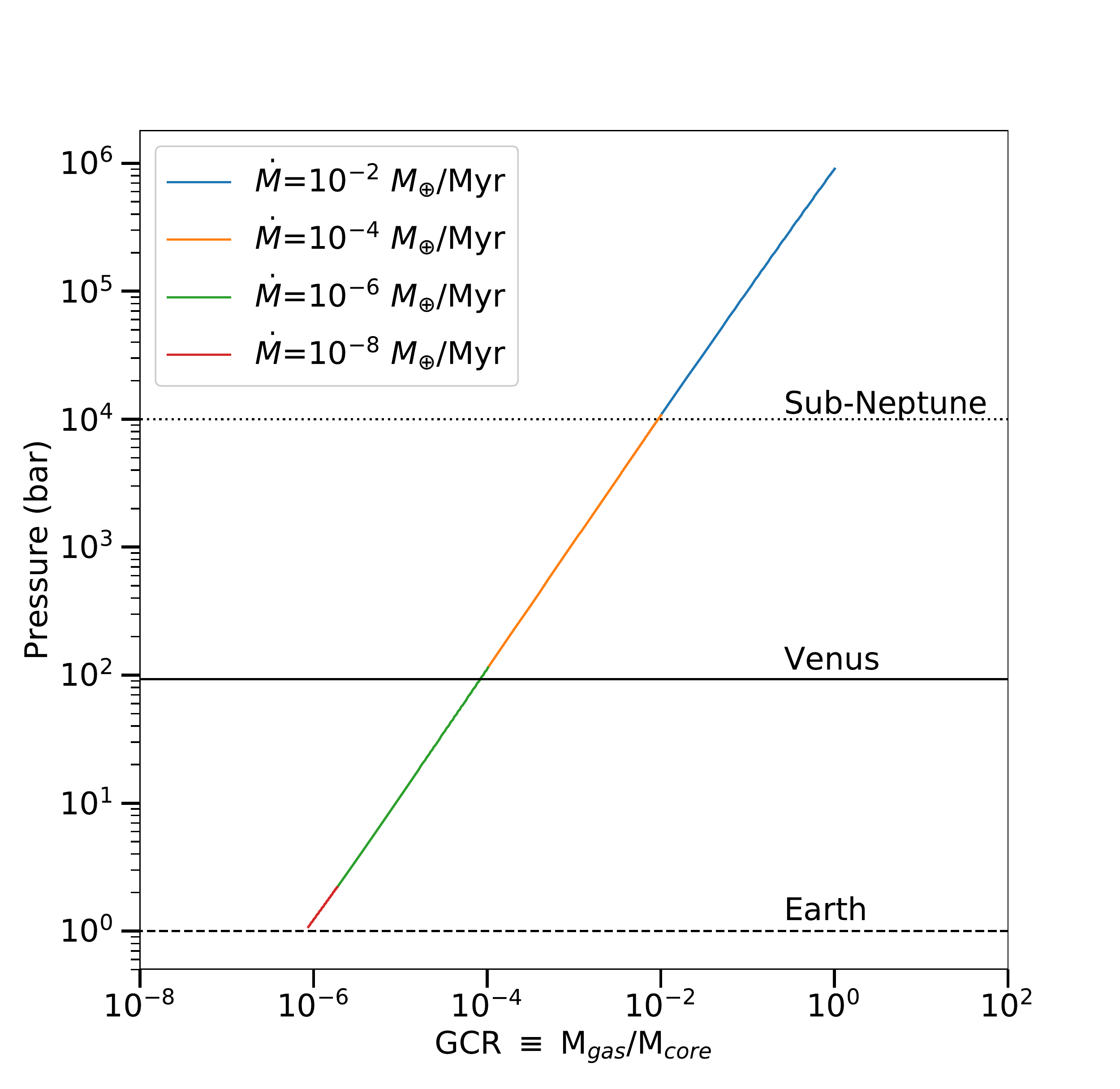}

\caption{Extended data - {\bf GCR and pressures for a planet starting with an Earth atmospheric mass.} Temporal evolution of the gas-to-core ratio (left) and pressure (right) starting from a pre-existing atmosphere with an Earth atmospheric mass.}
\label{figpre}

\end{figure}

This example shows that the final atmosphere will be dominated by volatiles delivered from late gas accretion for all cases with $\dot{M}>10^{-8}$ M$_\oplus$/Myr, i.e. in all cases with belts more massive than the Kuiper belt (see Extended data figure~\ref{figmdottime}), which is the least massive belt we know of so far. One could start with an even more massive atmosphere and just shift the lines up by the pre-existing mass or pressure to predict the final atmospheric masses and pressures. For instance, starting with a Venus-like atmosphere would still end up with an atmosphere being dominated by late-accretion after 100 Myr for $\dot{M}>10^{-6}$ M$_\oplus$/Myr.

\subsection{Formation of cavities owing to planets}

Our accretion models for both Earth-like and giant planets (Supplementary information) suggest that accretion onto planets is very efficient and in most cases, the rate of gas that can be accreted is higher than what is available (i.e. $\dot{M}$). For this reason, when the gas disk spreads inwards and crosses a planet's orbit, it may not be able to spread further in as all of the inflowing gas is accreted onto the planet that is being crossed. However, this depends on $R_H/H$. As shown in Extended Figure~\ref{fighill}, for small or distant planets, $R_H$ can be smaller than the disk scaleheight $H$ and a certain fraction of the gas will flow inwards. We note that for atmospheres with low $\mu$, high $\gamma$ or accreting for a very long time, the theoretical cooling accretion rate may become smaller than $\dot{M}$ and gas would flow inwards. We also note that the gas model used for low-mass planets is 1-D and thus assumes an axisymmetric gas flow but the gas flow geometry around the planet may be more complicated in 3-D\cite{2015MNRAS.447.3512O}, which may lead to gas flowing inwards anyway\cite{2019arXiv190702763B}. However, this is still unclear how much these complications affect the overall 1-D accretion rates\cite{2015ApJ...811..101F}, but recent 3-D simulations find gas accretion rates that are comparable to 1-D derivations\cite{2013ApJ...778...77D}. In the end, we expect accretion to be efficient and a gap in density should be seen in the gas distribution after crossing a planet, which would pinpoint the accreting planet location.

\subsection{Gas distribution with ALMA to infer the planet position.}

In Figure~\ref{figcav}, we used carbon observations as a good tracer for these cavities instead of CO. This is for two main reasons. First, carbon emission in band 8 seems to be a better tracer of this gas than CO in either band 6 or 7 according to models\cite{2017MNRAS.469..521K} and to the first few observations of neutral carbon in these disks\cite{2017ApJ...839L..14H,2018arXiv181108439K,2018ApJ...861...72C}. Second, carbon is always expected to spread in the inner region, given enough time, while CO photodissociates in about 100 yr in unshielded disks and remains colocated with the parent belt of planetesimals, implying that the CO cavity observed is then due to photodissociation rather than accreting planets. Only in the case of massive gas disks CO can be shielded by carbon\cite{2018arXiv181108439K} and may have time to viscously spread further inwards than the planetesimal belt. For the latter case, for systems where CO had time to spread in the inner region where planets are located, then CO cavities could also be used to infer planets but carbon is much more general as it is not subject to photodissociation.

We now explain the details of how we produced the synthetic ALMA image shown in Figure~\ref{figcav}. We first created a density profile for a late gas disk at steady state that scales as $r^{-1}$ inwards\cite{2018arXiv181108439K}. The center of the belt is at 50 au where the gas temperature is 20 K (scaling as $r^{-0.5}$). An input rate of $10^{-3}$ M$_\oplus$/Myr of CO has been chosen. We put the planet at 10 au and therefore impose that the gas density drops to zero within 10 au.

We then use the radiative transfer code RADMC-3D\cite{2012ascl.soft02015D} to obtain the emission of the carbon fine structure line at 492.16 GHz assuming that the inclination is 30 deg and position angle 45 deg. We then use the CASA software to simulate an ALMA observation at a resolution of 0.12" in band 8 for an hour on source (we use the C43-5 configuration together with the C43-2 to get more extended emission). Finally, we create a moment-0 image of the final cube and obtain Figure~\ref{figcav}. We see that the gas disk is well detected and the cavity within 10 au is well resolved.

\end{methods}

\appendix
\section{Supplementary Information}

\setcounter{figure}{0}

\subsection{Accretion onto giants and gas accumulation in their outer envelopes.}

For giant planets, we assume that all gas that flows through the planet is accreted. This assumption is valid in two cases. First, the scaleheight $H$ of late gas disks is found to be small compared to protoplanetary disks (because of lower temperatures and higher mean molecular weight) and a criteria that emerged over the last few years is that when the Hill radius of the planet is greater than $H$, the majority of the gas flowing in will be accreted by the planet\cite{2015MNRAS.452.3085Y,2016MNRAS.460.1243R}. We can check in Extended data Fig.~5 that $R_H>H$ is most likely the case for planets with masses greater than 10 M$_\oplus$ (or $>$ 0.03 M$_{\rm Jup}$), and we therefore expect all gas mass flowing through such a planet to be accreted. The disk scaleheight in 49 Ceti where we have the best constraints so far\cite{2017ApJ...839...86H} is lower than 0.04 $r$, where $r$ is the distance of the disk to the host star. This translates as a condition on the planet mass $M_{\rm pl}$ to have $R_H>H$, leading to $M_{\rm pl}>6 \times 10^{-5} M_\star$ or roughly that the planet mass is greater than a few Neptune masses. Another possibility is that if the viscous spreading originates in the magnetorotational instability as suggested recently\cite{2016MNRAS.461.1614K}, the instability could be quenched by a rather low magnetic field of the order of 1 $\mu$G (because in this case the waves propagating the instability in these low density disks do not anymore fit in the disk scaleheight), which would be the case in the close vicinity to the planet. This would mean that all gas would accumulate at the planet's location as it cannot spread inwards anymore (unless the planet is able to provide some viscosity to the gas disk by exciting density waves that transport angular momentum locally, which may also spread the disk for high enough surface densities\cite{2001ApJ...552..793G,2017ApJ...839..100F}), which would also justify our initial assumption.

When CO gas accretes onto a Sub-Neptune or more massive planet, it can increase the average mean molecular weight of the existing atmosphere. After accretion and mixing, we can calculate the new metallicity $Z$ of the atmosphere (assuming it is initially solar) as a function of time (plotted in Supplementary data Figure~\ref{figmet2}). To do so, we calculate 

\begin{equation}
Z= \frac{{\rm [O]/[H]+([O]/[H])}_{\rm solar}}{({\rm [O]/[H])}_{\rm solar}  },
\end{equation}

\noindent where $({\rm [O]/[H])}_{\rm solar}=4.898 \times 10^{-4}$ and [O]/[H]=$\dot{M} t/\mu_{\rm CO}/(2 M_{\rm atm} H_{\rm abun}/ \mu_{\rm proto})$, with $H_{\rm abun}=0.912$, $\mu_{CO}=28$ and $\mu_{\rm proto}=2.22$\cite{2003ApJ...591.1220L,2009ARA&A..47..481A} (see also the metallicity heat-maps in Figure~3 computed at 100 Myr).

\begin{figure}
\centering
    \includegraphics[width=11cm]{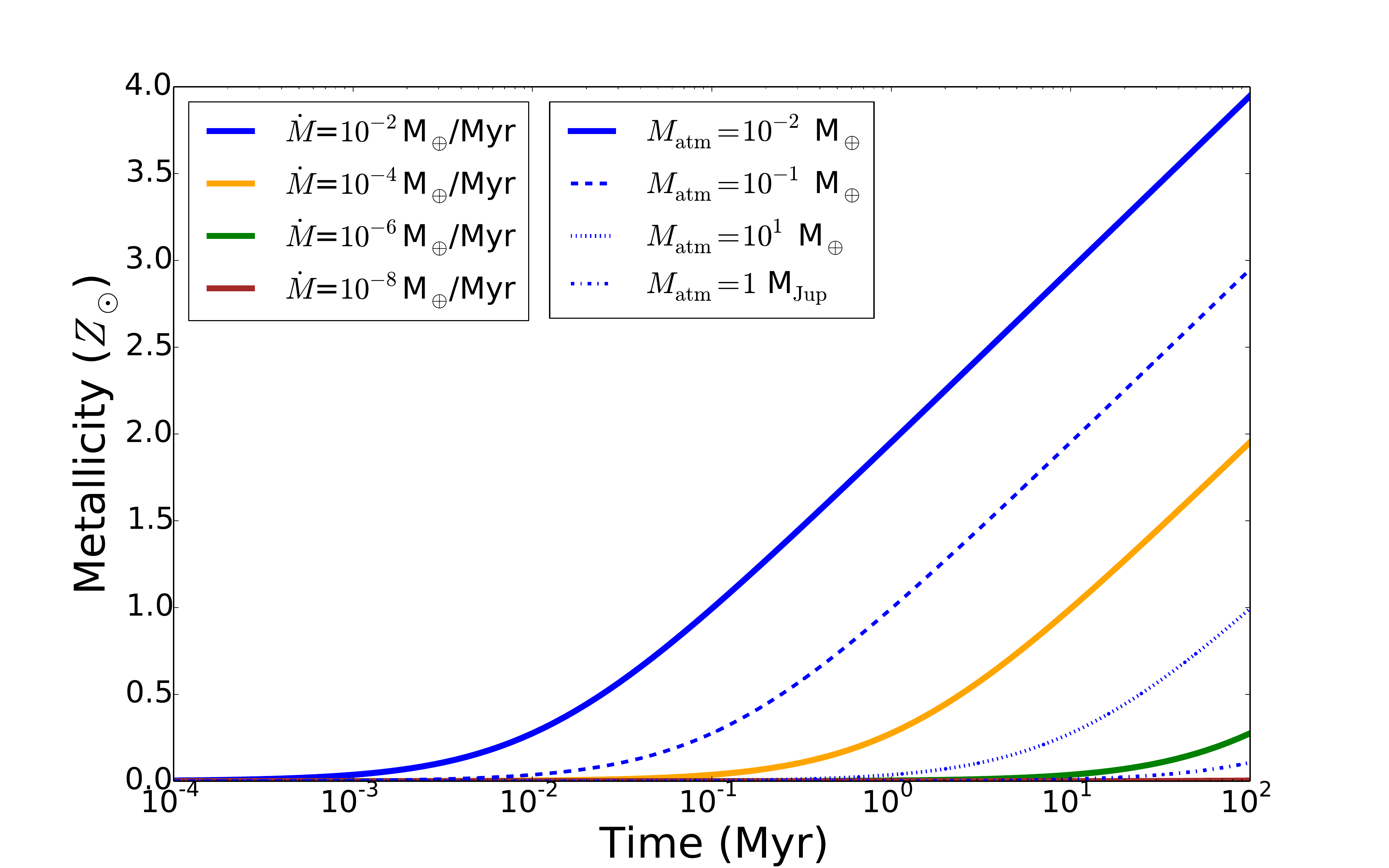}

\caption{Supplementary data - {\bf Evolution of metallicity for a planet accreting from a late gas disk.} Temporal evolution of the metallicity (in log) for different gas input rate and initial atmosphere masses.}
\label{figmet2}

\end{figure}

We also compute the evolution of the C/O ratio on the accreting planets assuming that the initial C/O ratio\cite{2009ARA&A..47..481A} is 0.5496 as follows

\begin{equation}
{\rm C/O}=\frac{0.5496+(Z-1)}{Z}, 
\end{equation}

\noindent where we fix $t$=100 Myr in Figure~3. This assumes that it is mostly CO that is released from the planetesimals but it could be that CO is trapped within CO$_2$ ices\cite{2019arXiv190709011S} and some CO$_2$ is also released and quickly photodissociates into CO, hence lowering the final C/O ratio to a value between 0.5 and 1 (if water is also released, which is not likely\cite{2016MNRAS.461..845K}, it would also lower the final C/O ratio). In Supplementary data Figure~\ref{figmet10Myr}, we also show our results after 10 Myr of evolution to see what can be expected in young systems.

\begin{figure}
\centering
    	\includegraphics[width=8cm]{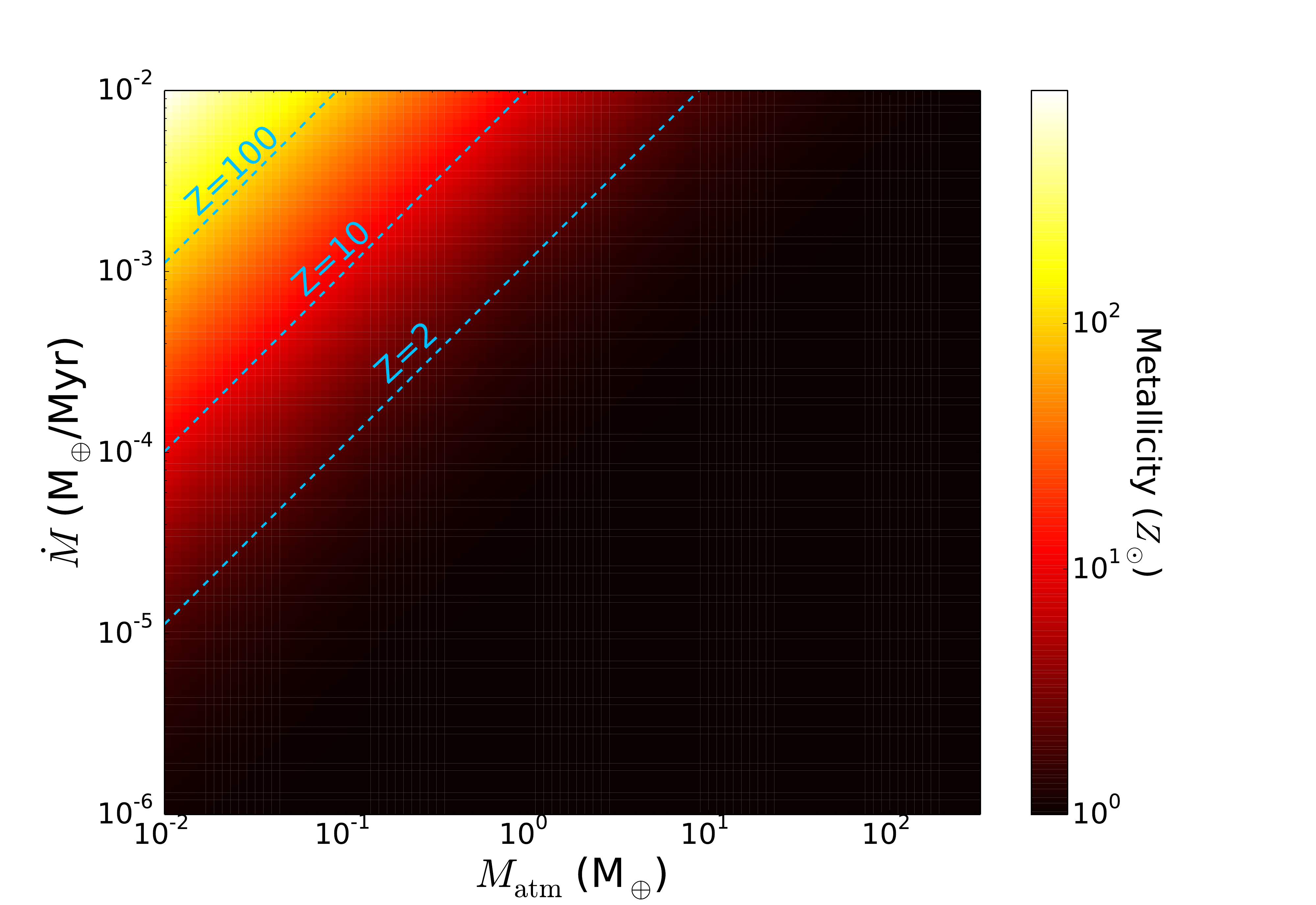}
        \includegraphics[width=8cm]{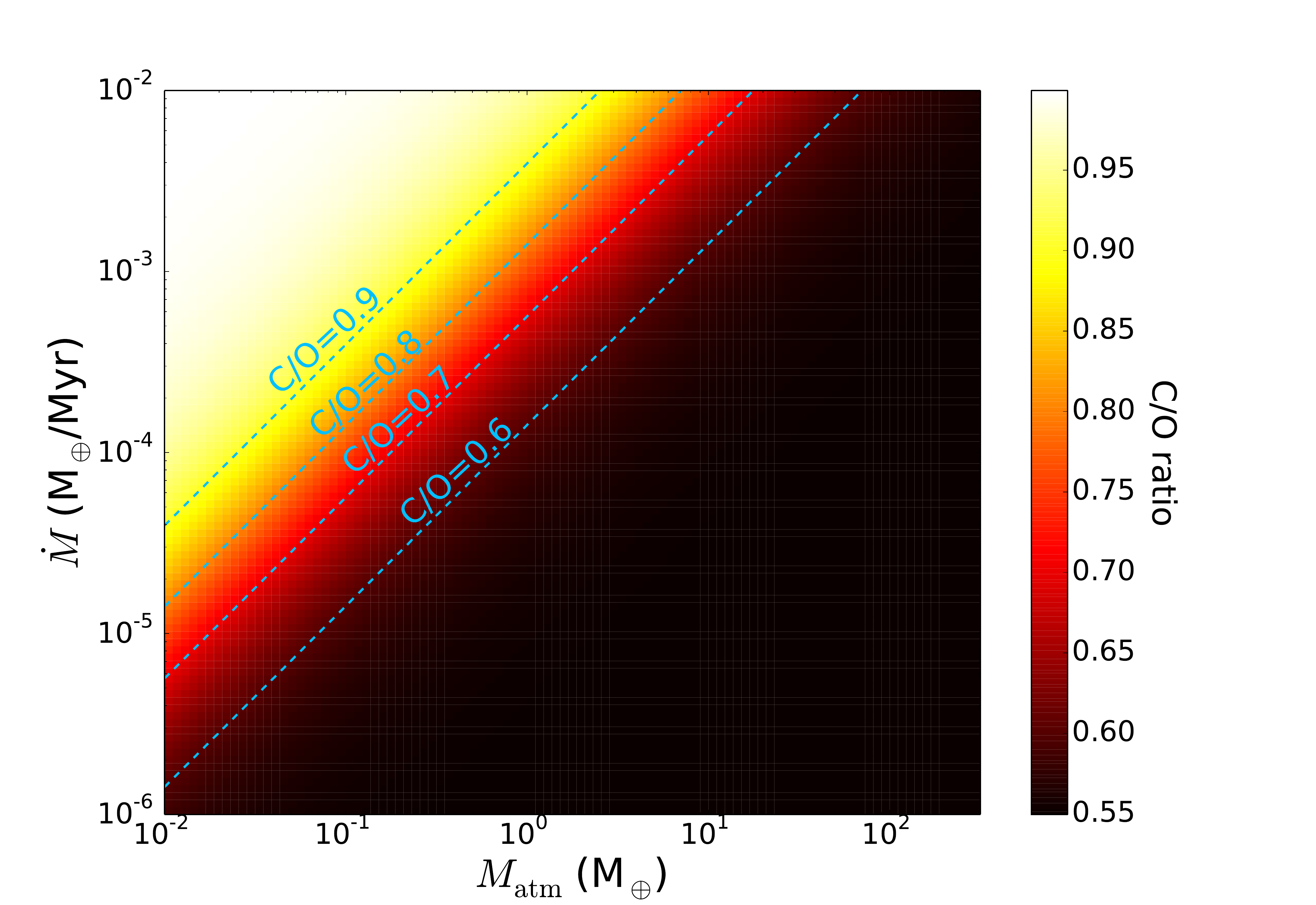}

\caption{Supplementary data - {\bf Signature of late gas accretion on young giant planets.} Temporal variation of metallicity (left) and C/O ratio (right) as accretion proceeds for 10 Myr (instead of 100 Myr in Figure~3) from an initially hydrogen-rich primordial atmosphere for different gas input rates and different initial atmosphere masses (Sub-Neptune up to Jupiter).}
\label{figmet10Myr}

\end{figure}

When material is accreted onto a Jupiter-like planet, gas first accretes in the outer envelope and then slowly diffuses inwards to a pressure $P$ on a timescale $t_{\rm diff}(P)$ that depends on the eddy mixing coefficient $K_{\rm zz}(P)$ such that\cite{2002Icar..159...95B}

\begin{equation}
t_{\rm diff}=\frac{2 H_{\rm atm}^2}{K_{\rm zz}(P)},
\end{equation}

\noindent where $H_{\rm atm}=k_b T/(\mu m_H g)$ is the atmospheric scaleheight, with $g=GM_p/R_p^2$ and $T$ is taken to be the skin temperature\cite{2010eapp.book.....S}. We will follow the gas that can accumulate in the upper enveloppe ($<$0.1 bar) and use the following prescription for $K_{\rm zz}$\cite{2018ApJ...854..172C}

\begin{equation}
K_{\rm zz}=10^{-2} \frac{H_{\rm atm}}{3}\left(\frac{R^\star \sigma T_{\rm eff}^4}{\mu \rho_{\rm atm} c_p}\right)^{1/3},
\end{equation}

\noindent where $R^\star$ is the universal gas constant, $\sigma$ the Stefan-Boltzman constant, $\rho_{\rm atm}$ the atmosphere density and $c_p$ the specific heat.

We then estimate the mass that accumulates at a pressure $P< 0.1$ bar (as most of the emission comes from the planet's photosphere close to 0.1 bar\cite{2018ApJ...854..172C}) as being $M_{\rm accu}=\dot{M} t_{\rm diff}$. Next, we estimate the gas mass that is already present $M_{\rm pres}$ at $<$ 0.1 bar assuming hydrostatic equilibrium so that $M_{\rm pres}=4 \pi R_{\rm pl}^2 P / g$. Finally, we focus on the CO mass that can accumulate at the surface compared to the CO mass that is already present. To calculate the latter, we use the mixing ratio of CO in the upper atmosphere derived from models, which varies with temperature as most of the carbon is in CO for warm planets (e.g. 1700 K) but in the form of CH$_4$ for the colder cases (e.g. 500 K). 

\begin{figure}
\centering
    \includegraphics[width=11cm]{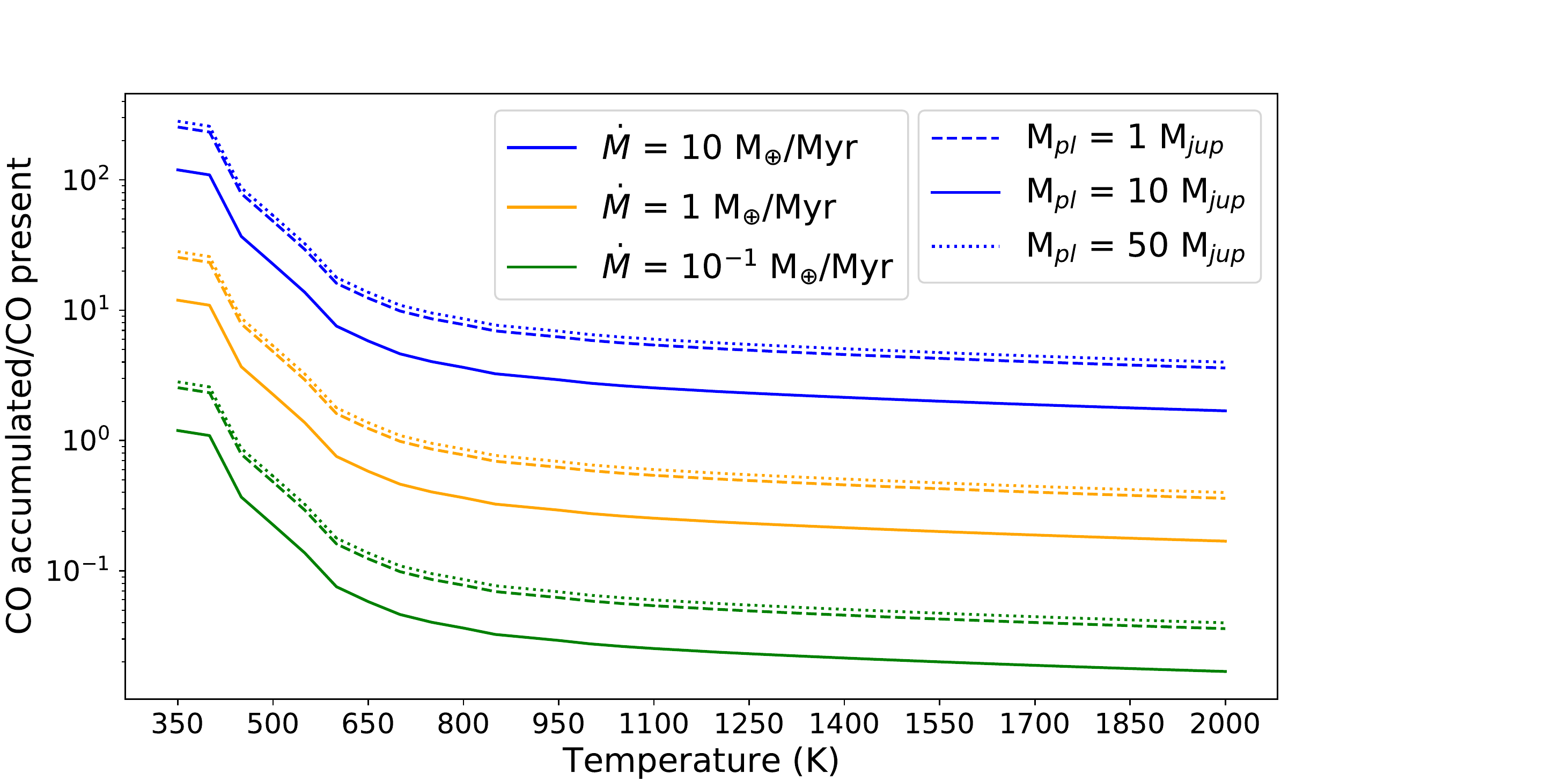}

\caption{Supplementary data - {\bf Gas accumulation in upper atmospheres of giant planets.} CO mass accumulated through accretion in a late gas disk over CO mass already present in a giant planet Vs. Temperature. The different colours show different input rates $\dot{M}$, while the solid, dashed and dotted lines are for a 10, 1 and 50 Jupiter-mass planet, respectively.}
\label{figpresent}

\end{figure}

Supplementary data figure~\ref{figpresent} shows the CO mass accumulated over the CO mass present calculated from our model as a function of the temperature for varying $\dot{M}$ (assuming all accreted gas is CO). As the diffusion timescale is often smaller than about a year, the gas input rate can be assumed much higher than when averaging over 100 Myr and it is reasonable to expect that bursts of CO ejection would lead to $\dot{M}$ values as large as 10 M$_\oplus$/Myr owing to e.g. a sudden change in viscosity at the planet location (due to a change in the magnetic field strength next to the planet or different ionisation fraction\cite{2016MNRAS.461.1614K}) for instance, or because when the system is very young and indeed very massive the non-averaged $\dot{M}$ can be very large (see Extended data Figure~2).
In Supplementary data Figure~\ref{figpresent}, we see that for the warm planet case, the amount of CO at the surface can only quadruple at most for a Jupiter-like planet. However, for a colder planet, as $K_{\rm zz}$ is smaller (hence it takes slightly longer for gas to diffuse inwards) and most carbon is originally in the form of CH$_4$ rather than CO (hence the contrast with the CO present is higher), the amount of CO at the surface can reach more than 100 times the amount of CO originally present for $T$=350 K. We note that the CO mass that can accumulate over the CO present do not vary with the mass of the planet but only with its radius (because $g$ cancels out in the ratio $M_{\rm accu}/M_{\rm pres}$ and this ratio scales as $R_{\rm pl}^{-2}$ because there is less CO present in smaller planets initially), and we find that more CO accumulates on Jupiter-like planets or brown dwarfs because their radii (calculated from mass and age\cite{2001RvMP...73..719B}) 1 and 0.95 $R_{\rm Jup}$, respectively, are smaller than that of a 10 M$_{\rm Jup}$ $\beta$ Pic-like planet equal to $\sim$ 1.46 $R_{\rm Jup}$.

\subsection{Detection of an accretion signature in giant planet spectra.}

\begin{figure}
\centering
    	                \includegraphics[width=8cm]{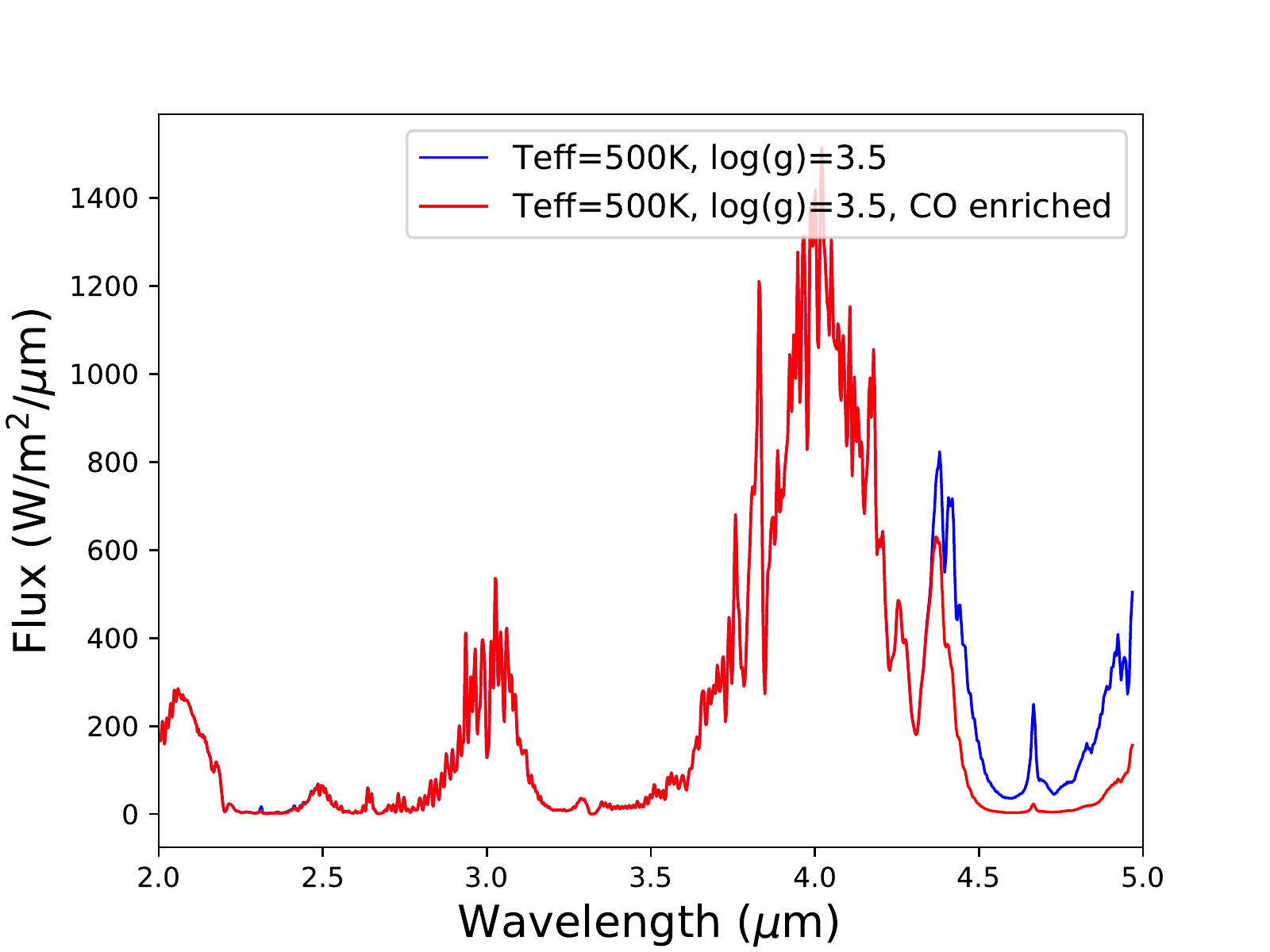}
	                \includegraphics[width=8cm]{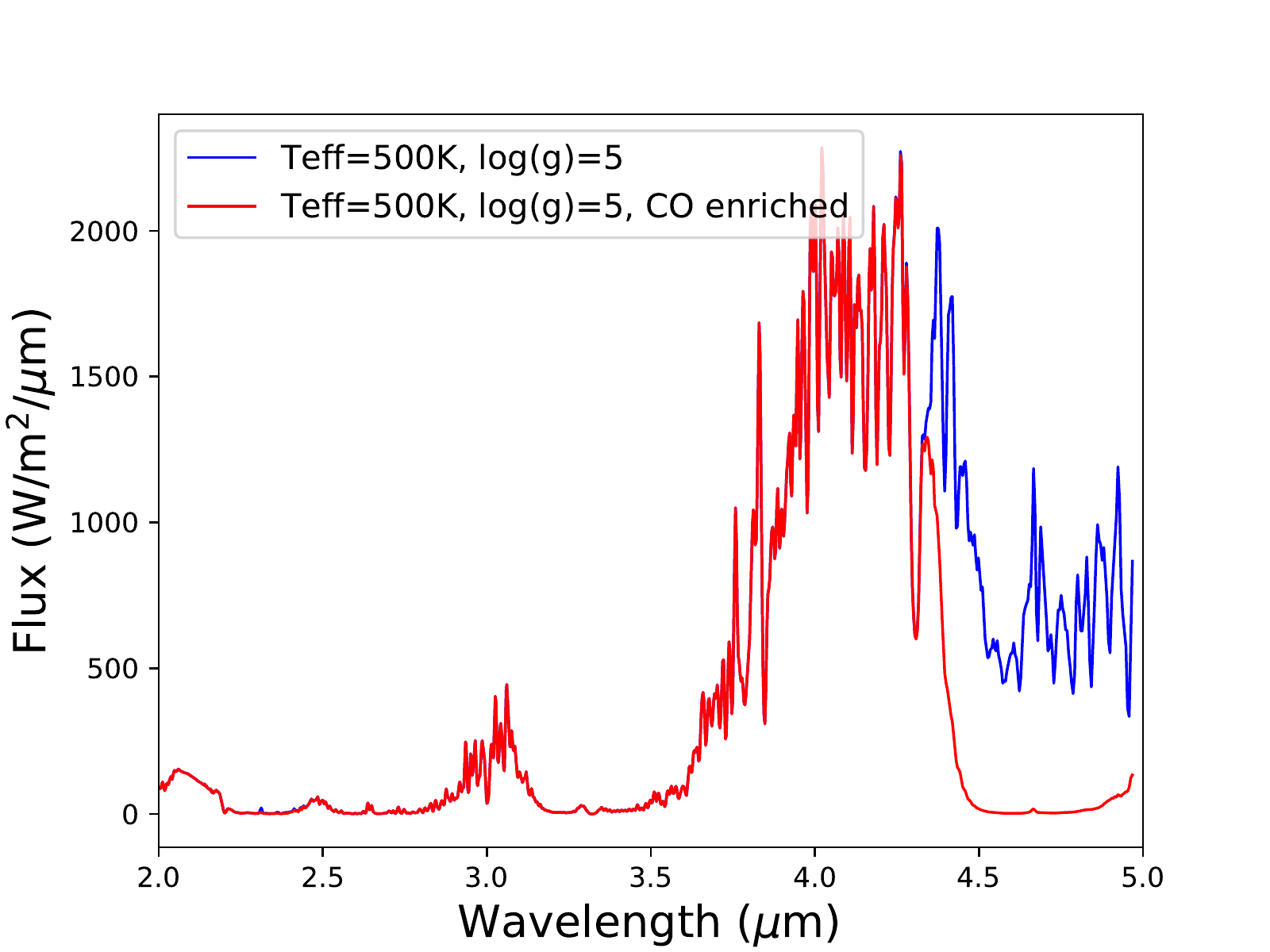}
    	                \includegraphics[width=8cm]{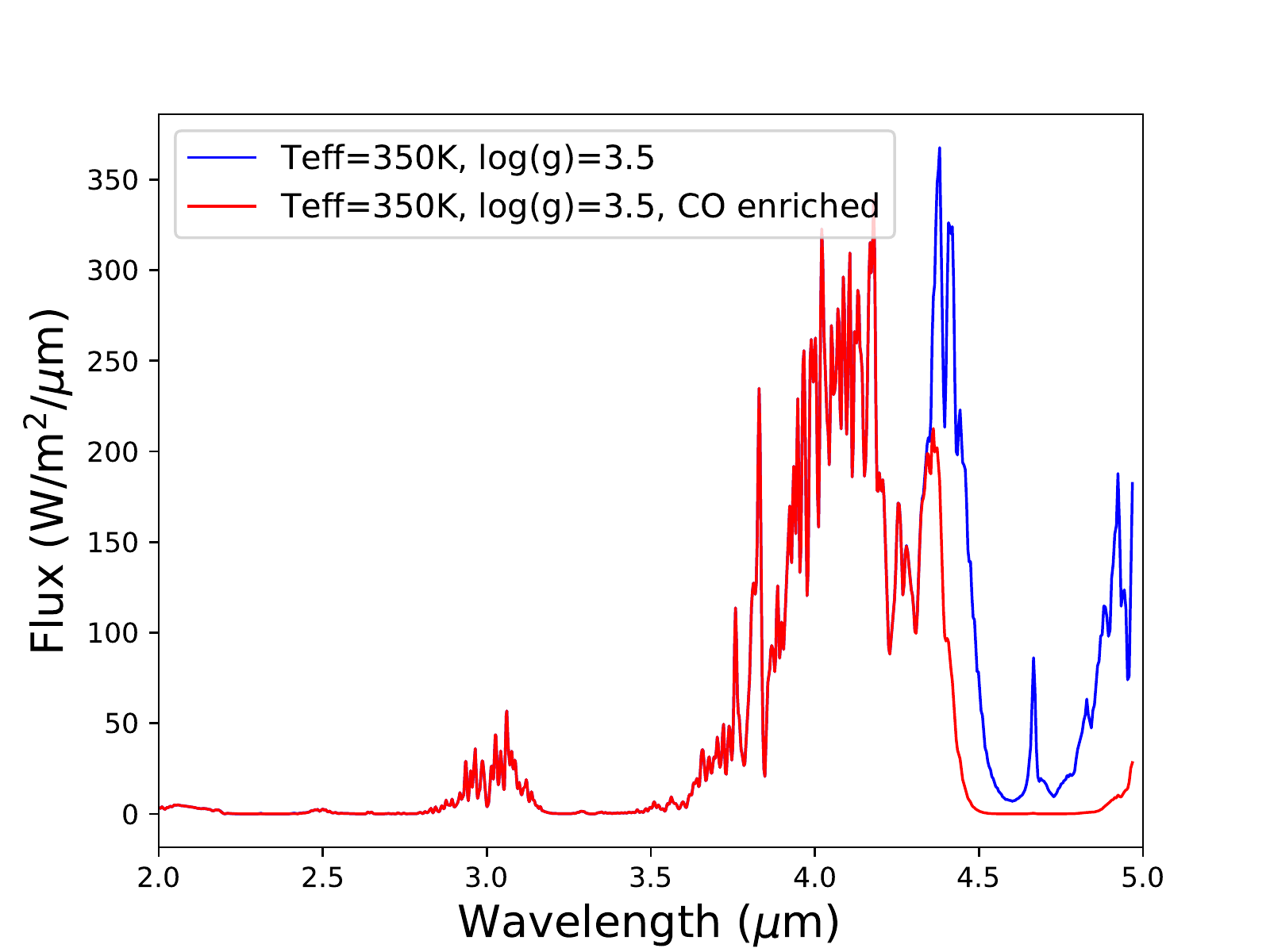}
	                \includegraphics[width=8cm]{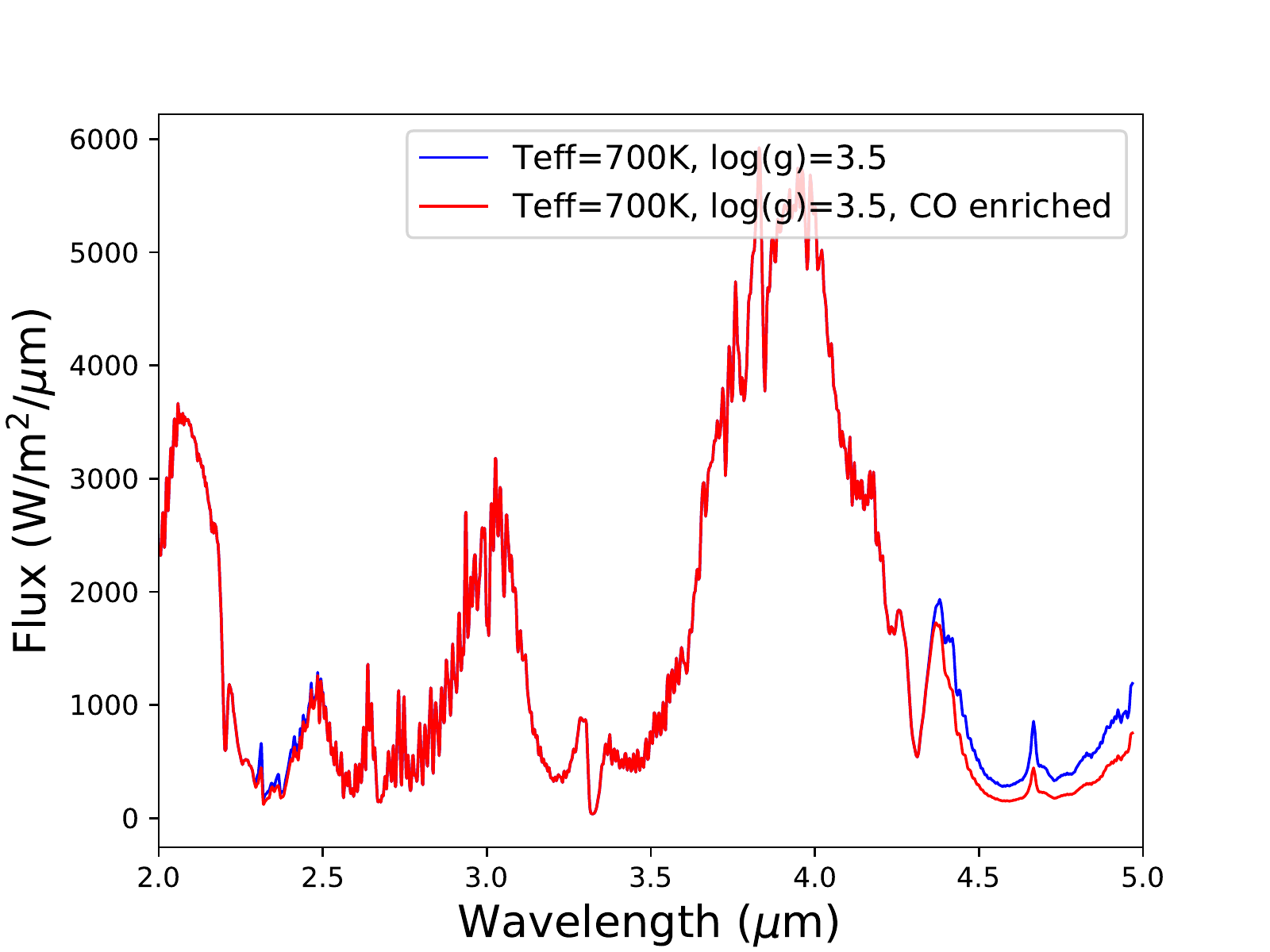}

\caption{Supplementary data - {\bf Signature of ongoing late gas accretion on giant planets.} Spectra of planets with different temperatures (350, 500 and 700 K) and masses ($\log(g)=3.5$ and $5$) suffering ongoing accretion of CO (in red) compared to the case with no late accretion (blue). In each case, we clearly see that the extra CO in the atmosphere creates a significant extra absorption in the M-band around 4.7 $\mu$m.}
\label{figspectre}

\end{figure}

We now estimate whether the excess of CO accumulated compared to CO present (see Supplementary data figure~\ref{figpresent}) can be detectable in atmosphere spectra. For a planet as hot as $\beta$ Pic b with a temperature of $T\sim$1700 K \cite{2015ApJ...815..108M}, the accumulation of gas in the outer layer of the atmosphere is not very efficient because the gas diffuses inwards very rapidly and maximising the gas input rate, we can only double (or quadruple for a 1 or 50 M$_{\rm Jup}$ planet) the amount of CO present in the outer layer ($<$0.1 bar) compared to the amount of CO expected in a primordial atmosphere with a solar metallicity. For a low temperature giant planet however, the inward diffusion is slower and assuming an $\dot{M}$ of 5 M$_\oplus$/Myr, up to 20 (5, 125) times more CO can be accumulated compared to the CO already present at $T=500$ K (700 K, 350 K) (see Supplementary data figure~\ref{figpresent}). This is mostly because in these low-temperature Jupiter-like planets, carbon is mostly in the form of CH$_4$ and the presence of so much CO would clearly be a sign of late accretion from late gas disks. Supplementary data figure~\ref{figspectre} shows the spectra of accreting planets with different masses ($\log(g)=3.5$ and $5$) and temperatures (350, 500 and 700 K) in red compared to their spectra with no ongoing accretion (in blue). We see that the extra accretion creates a significant extra absorption in the CO band between 4.5 and 5 microns that would be detectable with NIRSpec (IFU spectrograph) and NIRCam (imager with coronagraph) on the JWST and instruments on ELTs (e.g. IFU spectrograph HARMONI on the E-ELT) for the 500 and 700 K cases\cite{2018arXiv180303730B}. We see that the 500 K case is the most favourable and the brown dwarf case (with $\log(g)=5$) shows the biggest effect owing to the initially lower amount of CO in these atmospheres. Some planets that would be interesting to observe to confirm our scenario would be the 51 Eri b planet at 700 K (there is a JWST/NIRCam GTO observation planned) or GJ 504 b at 500 K.

The spectra shown on Supplementary data figure~\ref{figspectre} were computed using Exo-REM\cite{2015A&A...582A..83B,2018ApJ...854..172C} combined with a line-by-line radiative transfer code. We produced spectra ($R=6000$) using self-consistent silicate and iron clouds and non-equilibrium chemistry. Spectra between 2 and 5 microns are dominated by H$_2$O, CH$_4$ and CO absorptions.

\subsection{Maximum total mass available for accretion.}

Assuming that only CO is released from planetesimals in late gas disks (and no water or very little\cite{2016MNRAS.461..845K}), we can work out the maximum CO mass available that can potentially be accreted on a planet by estimating all the mass in CO initially available in the protoplanetary disk. The protoplanetary disk mass is roughly 0.5\% of the stellar mass\cite{2013ApJ...771..129A}. Assuming an A-type star similar to $\beta$ Pic, the total protoplanetary disk mass would be around 3000 M$_\oplus$. Assuming a standard CO-to-H$_2$ abundance ratio of $10^{-4}$, we find that the total CO mass would be around 4 M$_\oplus$. This is an upper limit of what could end up in the planetesimals of the belt as some of this CO gas will go into forming planets or will be lost through accretion/photoevaporation.

We conclude that a CO release rate of 0.1 M$_\oplus$/Myr is not ruled out in the early evolution of late gas disks (e.g. see Supplementary data figure~2) but would not be sustained for 100 Myr as all the initially available CO would be lost after a few tens of Myr at most. In this case where only CO is released, we cannot build planets as big as Saturn or Jupiter from an Earth-like core, unless water or other volatiles are abundant (but yet unseen) and released together with CO in these belts.

\subsection{Comparing to observations}

We have already presented a few key tests to corroborate our new mechanism to form atmospheres on terrestrial planets and Super-Earths. The most important of these would be observing the atmospheres of super-Earths and sub-Neptunes with JWST and ARIEL to look for atmospheres with high metallicities and high C/O ratios ($\sim$1) or trying to detect the signature of ongoing late accretion in very cool Jupiter-like planets. We have also shown that in contrast to the current planet formation paradigm of Super-Earths, we do not predict more massive cores to have higher GCRs, but rather the opposite, which needs further confirmation by observations. We now present a few more points that could be tested or that can help explaining current observations.

Our work suggests that the outermost planet of a system would be accreting most material. This means that given a core mass, the GCR of the outermost planet would be the highest and its density the smallest. Of course, planets within a same system do not always have the same core mass but we should see an overall trend that outermost planets (with $R_H/H>1$) must on average have lower densities. Though, this only works for planets that are not H$_2$-dominated for which the secondary late accretion of gas has a significant impact on the initial atmospheric mass (i.e., it is valid for desiccated planets after a giant impact or owing to photoevaporation and being at the bottom of the radius valley\cite{2017AJ....154..109F}). A recent study\cite{2016ApJ...817...90L} found that indeed there may be a trend of decreasing bulk density with increasing orbital period, which would need to be further studied in the future with e.g. new TESS results.

\subsection{Comparing to delivery from impacts}

The new scenario of late disk accretion we propose here is different from the late veneer or late accretion\cite{2015GMS...212...71M} proposed for the young Earth in which material was delivered through impacts rather than from gas disk accretion. It is thought that less than 1\% of an Earth mass of material was delivered by late impacts\cite{2016E&PSL.455...85B} to Earth. Therefore, assuming that the impactors are asteroid-like and have a maximum of 1\% of CO by mass\cite{2003SSRv..106..231G,2010Icar..208..438S}, we find that the total amount of CO that could have been delivered to Earth through impacts after the moon-forming collision is lower than $10^{-4}$ M$_\oplus$ (or GCR in CO of order $10^{-4}$). In our disk accretion scenario, the amount of CO delivered is at least equal to this for the least massive gas disks we considered, and can be orders of magnitude higher for more massive disks. One major difference between delivering volatiles through late disk accretion or impacts is that in the first scenario, most material delivered would likely be CO whereas in the impact scenario, water would dominate (and refractories), leading to a much lower C-to-O ratio for the latter.

Some impacts could also happen later in the system's history (after a few 100 Myr when late gas accretion may become much lower) due to instabilities perturbing an external belt and producing Large Heavy Bombardment-like (LHB-like) events\cite{2005Natur.435..466G} or owing to long term scattering of solids from an outer belt to the inner regions through a chain of planets\cite{2018MNRAS.479.1651M}. In our Solar System, it is found that accretion due to an effective source of comets such as a potential LHB leading to an accretion of $\sim 3\times 10^{-5}$ M$_\oplus$ of solids\cite{2005Natur.435..466G} would hardly change the Earth's atmospheric mass by more than 10\%\cite{wyatt+19}. However, the amount of mass accreted by a planet in this late stage could be higher in extrasolar systems and typically grow in mass by $\sim 1\%$ of the impactor mass accreted\cite{wyatt+19}. It is also known that if a certain mass of solids is scattered inwards from an outer belt, there is a 0.1-1\% accretion efficiency to reach planets in the terrestrial region\cite{2018MNRAS.479.1651M}. This could result in atmospheres $\sim$100 times more massive than that on Earth for bombardment involving $\sim 1$ to $10$ M$_\oplus$ of planetesimals scattered inwards, which is already more massive than the Kuiper belt by one to two orders of magnitude but could still be a fraction of a massive planetesimal belt. Assuming that 5\% of the belt mass is scattered over a few Gyr\cite{2018MNRAS.479.1651M}, it means that the total initial belt mass should be between 20 to 200 M$_\oplus$, which is very massive because in the most optimistic case we are left with 100 M$_\oplus$ of solid material at the end of the protoplanetary disk stage\cite{2011ARA&A..49...67W}. This means that for volatile delivery from very late impacts to be as important as what can be readily delivered from late gas accretion in low-mass disks, one requires the most top heavy belts, which are only a few in our surroundings and are negligible in number compared to the whole population. 

Moreover, there are also some limitations to the growth of an atmosphere from impacts. First, there is a specific region in the planet mass Vs. distance parameter space where an atmosphere can effectively grow and not deplete from impacts, which is defined by the line where the escape velocity becomes greater than the local Keplerian velocity of the planet\cite{2017MNRAS.464.3385W}. In contrast, in our scenario, all planets can grow. Second, there is a limit to atmospheric growth from impacts because when the atmosphere becomes too dense, atmospheric loss becomes more important and atmospheric growth might stall\cite{2014P&SS...98..120S}. Last but not least, the impact scenario requires impactors to be ejected and then to impact on the given planet (usually after a few interactions with other planets in the system). This is not very efficient and many comets/asteroids are ejected outwards or passed inwards without impacting the planets in a given system\cite{2018MNRAS.479.1651M}. In our scenario, once the gas is released, it will viscously diffuse inwards and will cross the planets' orbits, hence allowing accretion. This is why (in addition to stalling) atmospheres would never grow to reach Sub-Neptune pressure-like planets in an impact scenario\cite{2018MNRAS.479.2649K}, which they can however do from late disk accretion. 

Even worse, the scattering timescale of late impacts through scattering by a chain of planets can be very long. Generally speaking, the scattering timescale depends on the planet that is scattering solids inwards and on the chain of planets that will keep transferring the planetesimals inwards. It can be approximated by the cometary diffusion time\cite{2008Icar..196..274B} such that $t_{\rm scat} \sim M_\star^{3/2} a_{\rm pl}^{3/2} M_{\rm pl}^{-2}$, where $M_\star$ is the mass of the star in solar masses, $M_{\rm pl}$ the planet mass in Earth masses, $a_{\rm pl}$ the planet semi-major axis in au and $t_{\rm scat}$ in Gyr\cite{2018MNRAS.479.1651M}. This means that for planets with very low masses, the scattering timescale can be much longer than a Gyr and greater than the age of the system, therefore, making it impossible to scatter particles in at a high rate. For example, particles scattered by a 5 M$_\oplus$ planet at 50 au will evolve on timescales of $\sim 10$ Gyr, which is far too long for scattering to have an effect on the atmospheric compositions of impacted planets in most systems. One would need a much more massive planet that scatters solids inwards but if too massive (e.g. Jupiter-like), the planet is in the ejection regime\cite{2017MNRAS.464.3385W} and most particles are ejected, meaning that the scattering rate goes down. There is a strong compromise to find on the planet chain architecture for scattering to send solids in the inner regions. We conclude that our scenario is much more efficient than impacts and does not need any fine tuning of its planetary architecture or planet masses or positions for it to work. 

\subsection{Comparing to outgassing}

We can also evaluate our scenario in terms of volatile delivery compared to outgassing on a potential Earth-like planet. For a plate-tectonic degassing planet, we use the outgassing rate on Earth as an upper bound as plate tectonics is very active on our planet and we expect it to be similar or less efficient/active on other planets. Roughly 22 km$^3$ of basaltic magmas are produced each year on Earth\cite{1984JVGR...20..177C}. Therefore, we estimate the degassing rate\cite{2018MNRAS.479.2649K} to be $6 \times 10^{13}$ kg/yr (given the magma density of 2600 kg/m$^3$). Assuming a typical\cite{2001MinDe..36..490L} CO$_2$ content of 1wt\% and a perfect case of 100\% efficient degassing (with no recycling in the planet's mantle), we find an upper limit of $10^{-7}$ M$_\oplus$/Myr on the tectonically produced CO$_2$. This degassing rate is in the lower range of the CO input rate that the planet can accrete and can be orders of magnitude higher in our late gas disk accretion scenario. Degassing from volcanism would lead to a rather lower C-to-O-ratio (as it is mostly water and CO$_2$ that are released) than in our new proposed scenario.

\end{document}